# Twist-angle-dependent quantum phase diagrams in twisted bilayer $MoTe_2$

Zheng Sun[1†], Fan Xu[1,2†], Jiayi Li[1†], Yifan Jiang[3†], Jingjing Gao[4,8,9†], Cheng Xu[5,10†], Tongtong Jia[1,2], Kehao Cheng[1], Shaozheng Wang[1], Wanghao Tian[1], Jinyang Zhang[4], Kenji Watanabe[6], Takashi Taniguchi[7], Jinfeng Jia[1,2,8,9], Shengwei Jiang[1,2], Yang Zhang[5,11], Yuanbo Zhang[4,8,9,12*], Shiming Lei[3*], Xiaoxue Liu[1,2,8,9*], and Tingxin Li[1,2,8*]

[1]State Key Laboratory of Micro-nano Engineering Science, Key Laboratory of Artificial Structures and Quantum Control (Ministry of Education), School of Physics and Astronomy, Shanghai Jiao Tong University, Shanghai, China

[2]Tsung-Dao Lee Institute, Shanghai Jiao Tong University, Shanghai, China

[3]Department of Physics, Hong Kong University of Science and Technology, Clear Water Bay, Hong Kong SAR, China

[4]State Key Laboratory of Surface Physics and Department of Physics, Fudan University, Shanghai, China

[5]Department of Physics and Astronomy, University of Tennessee, Knoxville, TN, USA

[6]Research Center for Electronic and Optical Materials, National Institute for Materials Science, 1-1 Namiki, Tsukuba, Japan

[7]Research Center for Materials Nanoarchitectonics, National Institute for Materials Science, 1-1 Namiki, Tsukuba, Japan

[8]Hefei National Laboratory, Hefei, China

[9]Shanghai Research Center for Quantum Sciences, Shanghai, China

[10]Max Planck Institute for Chemical Physics of Solids, Dresden, Germany

[11]Min H. Kao Department of Electrical Engineering and Computer Science, University of Tennessee, Knoxville, Tennessee, USA

[12]New Cornerstone Science Laboratory, Fudan University, Shanghai, China

[†]These authors contribute equally to this work.

[*]Emails: zhyb@fudan.edu.cn, phslei@ust.hk, xxliu90@sjtu.edu.cn, txli89@sjtu.edu.cn

**Abstract:**

Two-dimensional semiconducting moiré superlattices provide a highly tunable platform for investigating strongly correlated and topological quantum phases [1-5]. As a prototypical example, twisted bilayer $MoTe_2$ ($tMoTe_2$) has been shown to host fractional topological phases at zero magnetic field [6-20]. However, their evolution with twist angle and competition with other correlated quantum states remain largely unexplored. Here we report a systematic transport study of twist-angle-dependent phase diagrams in $tMoTe_2$ across a range of 3.79°–5.76°, revealing an evolution from valley-polarized, fractional topological states to valley-degenerate correlated states and superconductivity. At relatively small twist angles, partially-filled Chern bands of $tMoTe_2$ host fractional quantum anomalous Hall (FQAH) states following the Jain sequence, together with signatures of an anomalous composite Fermi liquid at moiré hole filling factor $\nu_h = 1/2$. Increasing twist angle progressively suppresses fractional topological phases and reconstructs the half-filled Chern band into symmetry-breaking integer Chern insulating states. At $\nu_h = 1$, we observe a transition from robust integer quantum anomalous Hall (IQAH) insulators at small angles to displacement-field-tuned, topologically trivial correlated insulators at larger angles. Remarkably, at twist angles above approximately 5.6°, superconductivity emerges adjacent to the $\nu_h = 1$ correlated insulator, with a phase diagram closely resembling that recently reported in twisted bilayer $WSe_2$ [21-24]. Our results uncover a unified twist-angle-driven phase evolution linking fractional topology, symmetry breaking, magnetic order, and superconductivity, providing new insight into the emergent quantum phenomena in moiré systems.

**Main:**

Twisted semiconducting transition-metal dichalcogenide (TMD) homobilayers have emerged as a model system that is both theoretically tractable and experimentally rich. Within a continuum-model framework [25-28], the low-energy moiré valence bands at the *K*-valley of $tMoTe_2$ and twisted bilayer $WSe_2$ ($tWSe_2$) are predicted to be similar, both supporting topologically nontrivial moiré flat bands that effectively emulate the Kane–Mele model [29]. These topological flat bands are highly tunable: an applied vertical electric field can induce layer polarization and therefore drives a transition toward a topologically trivial regime. Furthermore, both the moiré band topology and the resulting interaction-driven quantum phases are predicted to depend sensitively on the twist angle $\theta$ [27,28,30-43].

Interestingly, experiments have revealed a clear distinction in the emergent phase diagrams of the two systems. In $tMoTe_2$, previous studies uncovered robust ferromagnetism together with integer and fractional Chern insulators (FCIs) over a broad twist-angle range of 2.1°–3.9° [6-18]. More recently, in high-quality $tMoTe_2$ devices around 3.8°-3.9°, reentrant integer quantum anomalous Hall (RIQAH) states and superconductivity are observed in close proximity to FQAH phases, where the superconducting state develops from a spin-valley-polarized Fermi surface exhibiting an anomalous Hall (AH) response [16]. In contrast, $tWSe_2$ exhibits zero-field integer Chern insulating states only within a narrow window below ~2° [44,45], although itinerant ferromagnetism persists up to ~2.7° [46]. And so far, no FCIs or other fractional topological phases has been observed in $tWSe_2$. Instead, robust superconductivity has been observed in $tWSe_2$ at larger twist angles between approximately 3.5° and 5° [21-24], arising from a valley-degenerate Fermi surface and are considered to be associated with intervalley-coherent antiferromagnetic (AFM) order [47-51]. The origin of these striking differences between otherwise theoretically analogous moiré TMD systems remains unclear. And a systematic experimental study of how fractional topology, magnetic order, and emergent superconductivity evolve across $\theta$ has remained lacking.

Here we address this question by systematically mapping the transport phase diagram of $tMoTe_2$ across fifteen devices (see Extended Data Table 1 for device summary), covering a twist-angle range of 3.79°–5.76°. The twist angle $\theta$ is calibrated through quantum oscillations under magnetic fields (Extended Data Fig. 1). Most devices were fabricated using high-quality, flux-grown 2H-$MoTe_2$ single crystals (Methods).

**Twist-angle-dependent phase diagram in $tMoTe_2$**

Figure 1a shows a schematic of the $tMoTe_2$ device used for transport measurements (Methods). Figure 1b presents the calculated density of states (DOS) as functions of interlayer potential difference $\Delta$ and $\nu_h$, for 5.8° $tMoTe_2$ as an example (Methods). At zero electric field, the van Hove singularity (VHS) lies below $\nu_h = 1$, and it shifts to higher $\nu_h$ with increasing $\Delta$. By varying $\theta$, the moiré wavelength and the resulting electronic bandwidth can be continuously tuned, as shown in Fig. 1c. Figures 1d-1f and 1j-1l present the longitudinal resistivity $\rho_{xx}$ as functions of $\nu_h$ and vertical displacement electric field $D$, measured under a small perpendicular magnetic field $B$ (or at $B = 0$) for six representative $tMoTe_2$ devices (also see Fig. 3, and Extended Data Fig. 2 for other devices). For devices with $\theta$ smaller than approximately 4°, the $\rho_{xx}$ (Fig. 1d and Extended Data Fig. 2a) and Hall resistivity $\rho_{xy}$ (Fig. 1g and Extended Data Fig. 2b) maps closely resemble those reported previously [8,9,13-16]. Specifically, in the layer-hybridized regime, vanishing $\rho_{xx}$ around $D = 0$ signals the formation of interaction-induced valley-polarized Chern insulating states, including the IQAH state at $\nu_h = 1$, and FQAH states at $\nu_h = 2/3$, 3/5, and 4/7. In the layer-polarized regime, pronounced insulating states driven by strong electron-electron interactions emerge at $\nu_h = 1$, 2/3, 1/2, 1/3, and 1/4. These states are topologically trivial, and can be understood as a Mott insulating state ($\nu_h = 1$) and generalized Wigner crystal states on the triangular moiré superlattice [1,2].

As the $\theta$ increases, both IQAH and FQAH states are gradually suppressed. The corresponding $\rho_{xx}$ minima weaken and eventually disappear, reflecting a gradual destabilization of chiral edge transport. For devices with $\theta > 5°$, no signatures of magnetic hysteresis are detected at any fillings, demonstrating the fully suppression of spontaneous valley polarization. Instead, correlated insulating features remain in the layer-hybridized regime, particularly at $\nu_h = 1$. Taken together, these observations demonstrate a systematic evolution from a strongly interacting, valley-polarized topological regime at small angles to valley-degenerate regime at larger angles.

The weakening of electronic interactions with increasing $\theta$ is further evidenced by the evolution of topologically trivial correlated insulating states in the layer-polarized regime. With increasing $\theta$, the layer-polarized correlated insulating states become significantly weakened, consistent with an increased bandwidth and a reduced interaction-to-kinetic-energy ratio. Phenomenologically, we observe that these layer-polarized correlated insulating states disappear sequentially with increasing $\theta$, following the order of $\nu_h$ = 1/4, 1/3, 2/3, 1/2, and finally 1. In addition, for small-angle devices and near $D = 0$, a broad insulating region emerges that is continuous in filling factor: for $\theta < \sim 5.2°$, the system remains insulating for roughly $\nu_h < 0.4$ around $D = 0$. As the $\theta$ increases, this continuous insulating regime progressively shrinks. By 5.76°, only discrete insulating states at $\nu_h$ = 1/4, 1/3 remain, with the intervening fillings becoming metallic (Fig. 1l).

**The evolution of FQAH and ACFL phases with $\theta$**

Next, we investigate the evolution of fractional topological phases in partially-filled Chern bands with $\theta$. Figure 2a-2e show the symmetrized $\rho_{xx}$ and antisymmetrized $\rho_{xy}$ as a function of $\nu_h$ at $D = 0$ and $B = 0.1$ T for device 1-4 and 6, respectively. For device 1 with $\theta = 3.79°$, three prominent FQAH states are observed at $\nu_h$ = 2/3, 3/5, and 4/7, following the Jain sequence. These states exhibit pronounced local minima in $\rho_{xx}$ accompanied by quantized anomalous Hall plateaus in $\rho_{xy}$ (Fig. 2a), consistent with previous studies [8,9,13-16]. Notably, at $\nu_h$ = 1/2, the AH resistivity approaches $2h/e^2$ but without a plateau feature in $\rho_{xy}$ and a local minimum in $\rho_{xx}$, demonstrating a compressible nature of the state. Theoretically, a compressible state with AH response close to $2h/e^2$ at half-filled Chern band, together with the emergence of FQAH states following Jain-sequence at partially-filled Chern band constitutes a hallmark of an anomalous composite Fermi liquid (ACFL) rather than a conventional Fermi liquid [31,52-54]. Recent optical measurements have provided further support for the existence of such ACFL phases [18].

As the $\theta$ increases, the number of observable FQAH states gradually decreases. At $\theta$ = 4.08° (device 3), only the $\nu_h$ = 2/3 FQAH state remains, and its $\rho_{xy}$ already significantly deviate from the quantized value of $3h/2e^2$ (Fig. 2c). Upon further increasing the $\theta$ to 4.14°, all FQAH states disappear, although ferromagnetic order still persists over the filling range of $\nu_h \approx 1/2$ to 1 at $D = 0$ (see Extended Data Fig. 3). Meanwhile, the AH resistivity at $\nu_h$ = 1/2 drops below $2h/e^2$ with increasing $\theta$, demonstrating the destabilization of the ACFL phase. Notably, in devices with $\theta$ between 4.14° and 4.69°, integer Chern insulating states emerge at fractional $\nu_h$, spontaneously breaking the translational symmetry of the underlying moiré superlattice. Specifically, in device 4 ($\theta = 4.14°$), we observe an integer-quantized AH plateau of $h/e^2$ in $\rho_{xy}$, accompanied by a local minimum in $\rho_{xx}$, centered at $\nu_h \approx 0.62$ (Fig. 2d). For devices 6-8 ($\theta$ = 4.55°-4.69°), integer Chern insulating states appear at $\nu_h$ = 1/2, as shown in Fig. 2e and Extended Data Fig. 4, 5. These observations suggest a twist-angle-driven topological phase transition at half-filled Chern band, in which the compressible ACFL gives way to symmetry-broken integer Chern insulating (SBCI) phases, consistent with recent theoretical calculations [35]. Figure 2f summarize the evolution of the AH resistivity with $\theta$ at $\nu_h$ = 2/3 and 1/2. Both signals collapse rapidly once $\theta$ exceeds ~4.0°, marking the disappearance of fractional topological phases, while an intermediate SBCI state emerges at $\nu_h$ = 1/2 before the valley polarization is fully suppressed.

## IQAH insulator, correlated insulator and superconductivity at $\nu_h = 1$

We next focus on the twist-angle evolution of correlated and topological phases at $\nu_h$ = 1. As shown in Fig. 1 and Extended Data Fig. 2, for devices with $\theta < 4.22°$ (device 1-5), robust IQAH states are observed over a continuous range of $D$-field around $D = 0$, while topologically trivial correlated insulating states appear at larger $D$. Interestingly, in device 6-8 ($\theta$ = 4.55°-4.69°), we noticed that the IQAH state at $\nu_h$ = 1 no longer exists at $D$ = 0 but instead emerges only at finite $D$, as shown in Figs. 1f, lj, 3a, 3b and Extended Data Fig. 2g, 2h. Notably, the emergence of IQAH states roughly coincides with the $D$-field at which the single-particle VHS crosses $\nu_h$ = 1 (Extended Data Fig. 6). The Curie temperature for the ferromagnetism at $\nu_h$ = 1 in device 6-8 is approximately 3-4 K (Extended Data Fig. 4), significantly lower than the values previously observed at smaller $\theta$ (~10-15 K in 3.5°-4° devices) [8,9,15]. Such reduction reflects the weakening of valley polarization with increasing $\theta$. Figure 3c shows the temperature-dependent $\rho_{xx}$ as a function of $D$ at $\nu_h$ = 1 in device 7 (4.63°). Within the IQAH regime, the high-temperature transport behavior is metallic, followed by an upturn in $\rho_{xx}$ upon cooling below approximately 6 K; upon at temperatures below 4 K, the $\rho_{xx}$ decreases as edge-state transport becomes dominant. An Arrhenius fit to this low-temperature decrease yields a gap of approximately 1 meV for the IQAH state (Fig. 3f).

With further increasing $\theta$ to around 5°, both the IQAH state at $\nu_h$ = 1 and ferromagnetism at other fillings are completely suppressed. Figure 3d shows $\rho_{xx}$ as functions of $\nu_h$ and $D$ for device 10 (4.99°). Although $\rho_{xx}$ features within the layer-hybridized region remain qualitatively similar to those observed in device 7 (4.63°), a topologically trivial correlated insulating state appears instead of the IQAH insulator at comparable $D$ values where the VHS crosses $\nu_h$ = 1. The activation gap of the correlated insulating state, extracted from the $T$-dependent $\rho_{xx}$ (Fig. 3e), is approximately 2.5 meV.

Figure 3g and 3h displays $\rho_{xx}$ and $\rho_{xx}$, respectively, as functions of $v$ and $D$ in device 15 (5.76°), measured at a nominal base $T$ = 10 mK (Methods). At this larger twist angle, correlation effects are further weakened. Aside from the correlated insulating states at $v_h$ = 1/4 and 1/3 near $D$ = 0, together with the $v_h$ = 1 insulating state and nearby features at finite $D$, most remaining transport features can be well captured by a single-particle band framework (Extended Data Fig. 6). Remarkably, superconductivity emerges in close proximity to the correlated insulating state at $v_h$ = 1, as detailed in the next section.

Theoretical calculations [37-41] have suggested that at $v_h$ = 1, twisted TMD homobilayers host a hierarchy of competing interaction-driven phases with diverse magnetic orders, whose relative stability can be tuned by $\theta$ and $D$. In the layer-hybridized regime, calculations predict a multiferroic phase with spontaneous layer polarization and magnetism at very small $\theta$, a valley-polarized IQAH insulator with Chern number $C$ = 1 at intermediate $\theta$, and topologically trivial AFM orders at larger $\theta$, where increased bandwidth and reduced topological character suppress valley polarization. The evolution of the latter two regimes with $\theta$ is in qualitative agreement with our experimental observations, as summarized in the experimental phase diagram shown in Fig. 3i. Moreover, the fact that insulating states at $v_h$ = 1 in $\theta$ > 4.5° $tMoTe_2$ devices emerge only when the VHS approaches filling $v_h$ = 1 illustrates that the enhanced DOS associated with the VHS amplifies interaction effects and promotes fully-gapped insulating phases, either topologically nontrivial or trivial, over metallic phases.

## Superconductivity in large-angle $tMoTe_2$

Figure 4 further illustrates the characteristics of the superconducting states emerging near $v_h$ = 1 in 5.76° $tMoTe_2$. The finely scanned local map (Fig. 4a) reveals that superconductivity develops adjacent to the low $|D|$ end of the $v_h$ = 1 correlated insulator and becomes most robust at fillings slightly below $v_h$ = 1 (Extended Data Fig. 7). Figure 4b shows the $\rho_{xx}$ as functions of $D$ and $T$ at $v_h$ = 1, where insulating, superconducting, and metallic behaviors appear sequentially with decreasing $D$ in the layer-hybridized regime. Representative $\rho_{xx}$-$T$ curves shown in Extended Data Fig. 8 illustrate the transport characteristic for these three regions. In the insulating region, an energy gap of approximately 2.5 meV is extracted. In the superconducting region, the normal state exhibits strange-metal behavior, whereas the metallic region at smaller $|D|$ follows Fermi-liquid behavior.

Figure 4c presents the $\rho_{xx}$ as functions of $v_h$ and $T$ cut across the optimal superconducting region, with the optimal superconducting transition temperature $T_c$ about 225 mK (determined by 50% normal resistivity $R_n$). This is further evident by the measured differential resistivity d$V$/d$I$ as functions of the dc bias current $I_{dc}$ (Fig. 4d).

Figure 4e shows d$V$/d$I$ as functions of $B$ at $T$ = 10 mK, from which a perpendicular critical magnetic field $B_c \approx 100$ mT can be identified. Figure 4f shows the $\rho_{xx}$-$T$ curves measured under different $B$ at optimal $\nu_h$ and $D$ for the superconducting state. From a Ginzburg–Landau analysis of the temperature dependence of $B_c$ (inset of Fig. 4f), the superconducting coherence length is estimated to be $\xi \approx 44$ nm (Methods). The 5.76° twist angle corresponds to a moiré wavelength of $a_M \approx 3.5$ nm, and the mean free path in this device is estimated to be $l_m \approx 120$ nm (Methods). These values suggest tightly bonded Cooper pairs ($\xi/a_M \approx 12.5$), and the superconductor is in the clean limit ($\xi/l_m \approx 0.37$). Superconducting states near the $\nu_h$ = 1 correlated insulator are also observed in device 13 (5.67°) and 14 (5.69°), with slightly lower $T_c$, as shown in Extended Data Fig. 9.

Phenomenologically, the observed superconducting characteristic and phase diagram in large-angle $tMoTe_2$ ($\theta > 5.5°$) resemble the recent reported superconductivity in $tWSe_2$, especially for those devices around 4.0°-4.3° [23,24]. Clear Fermi surface reconstructions are also observed in device 15 (Extended Data Fig. 10), similar to those reported in the ~4°–5° $tWSe_2$ devices [22-24]. Theoretical studies have associated superconductivity in $tWSe_2$ with the strongly correlated electronic states near such Fermi-surface reconstructions, highlighting intervalley fluctuations, quantum criticality, and band topology as possible ingredients (see Methods). The observation of analogous superconductivity in large-angle $tMoTe_2$ therefore provides an important opportunity to clarify the microscopic origin of this class of moiré superconductors through direct comparison between different material platforms.

## Discussions and conclusions

In this work, we address how $tMoTe_2$ evolves from a regime dominated by valley-polarized fractional topological order to one supporting valley-degenerate superconductivity. At relatively large $\theta$, the transport phase diagrams of $tMoTe_2$ and $tWSe_2$ become remarkably similar, with valley-degenerate superconductivity emerging near the $\nu_h$ = 1 correlated insulating state within the layer-hybridized regime. In general, $tMoTe_2$ exhibits stronger correlation effects at comparable moiré wavelengths, likely due to the larger effective mass of monolayer $MoTe_2$ (~ 0.75 $m_0$) [55] compared with that of monolayer $WSe_2$ (~ 0.4 $m_0$) [56]. For example, although 5.0° $tWSe_2$ ($a_M \approx 3.8$ nm) and 5.76° $tMoTe_2$ ($a_M \approx 3.5$ nm) possess similar moiré wavelength, correlated insulators are absent in 5.0° $tWSe_2$ but remain robust in 5.76° $tMoTe_2$. On the other hand, the superconducting transition temperature is higher in 5.0° $tWSe_2$.

As the $\theta$ decreases and electronic correlations strengthen in both systems, their phase evolution diverges significantly. In $tWSe_2$, ferromagnetism and QAH states are difficult to stabilize; even at 2.7°, only trivial correlated insulating behavior has been reported at $\nu_h = 1$ and $B = 0$ [23,46]. By contrast, 4.69° $tMoTe_2$ already hosts IQAH states at $\nu_h = 1$ and SBCI states at $\nu_h = 1/2$. In $tMoTe_2$ devices with $\theta < 4.08°$, fractional topological phases including FQAH and ACFL states further emerge. The competition among fractional topological ground states across $\theta$ might be understood from a quantum-geometric perspective: although correlations at larger angles can remain sufficiently strong to induce valley polarization in $tMoTe_2$ at fractional moiré fillings, the increased bandwidth and larger deviations from ideal quantum geometry may destabilize FCI and ACFL states relative to other competing states [31-35].

Our results suggest that superconductivity in $tMoTe_2$ may fall into two distinct classes. At small angles, as reported previously [16], unconventional superconductivity coexists with fractional topological phases and emerges from a spin-valley-polarized normal state. At large angles, as demonstrated in current work, valley polarization is fully suppressed and superconductivity instead develops from a valley-degenerate Fermi surface near $\nu_h = 1$, while no superconductivity has been observed at intermediate angles thus far (see Methods for more discussions). A systematic understanding of how superconductivity, and its underlying pairing mechanism, evolves with $\theta$ in $tMoTe_2$ therefore remains an important open question. Our results uncover a rich twist-angle–dependent landscape of quantum phases in $tMoTe_2$, and provide a fascinating platform for investigating how unconventional superconductivity emerges from competing correlated and topological electronic states.

## Methods

### 2H-$MoTe_2$ Crystal growth

High-quality 2H-$MoTe_2$ single crystals were synthesized independently at The Hong Kong University of Science and Technology (HKUST) and Fudan University (FDU). Although the detailed growth parameters differ slightly, both approaches follow similar tellurium-rich flux growth procedures. The growth recipes used at HKUST and FDU are described below.

**HKUST:** High-quality $MoTe_2$ single crystals were synthesized using an optimized vertical self-flux method. Molybdenum powder (99.997% Mo) and excess tellurium blocks (99.99999% Te) were thoroughly mixed at a molar ratio of Mo:Te = 1:110 and loaded into a quartz ampule. The ampule was evacuated to $\sim 10^{-5}$ Torr, subjected to multiple purge–pump cycles, and flame-sealed under vacuum. The sealed ampule was heated to 880 °C over 24 h, held for 12 h, and then slow-cooled at 1 °C/h to 530 °C. Crystals were separated from the molten flux by centrifugation at 530 °C. A subsequent post-annealing step was performed for 48 hours in a temperature gradient ($T_{\mathrm{hot}}$ = 400 °C, $\Delta T \approx 200$ °C) to remove residual tellurium and reduce tellurium inclusions, yielding high-quality $MoTe_2$ crystals suitable for mechanical exfoliation and 2D device fabrication.

**FDU:** High-purity molybdenum powders (99.997% Mo) and excess tellurium blocks (99.99999%) were thoroughly mixed at a molar ratio of 1:30 and loaded into a quartz ampule. The ampule was subsequently sealed under high vacuum (~$10^{-5}$ Torr). The sealed ampule was then heated to 1000 °C over 40 hours, followed by a 168-hours dwell at 1000 °C. Then, the ampule was cooled at a rate of 1 °C/hour to 550 °C before centrifuging. To further remove the excess tellurium flux, the $MoTe_2$ crystals were transferred to a second vacuum-sealed ampule and subjected to a temperature-gradient ($T_{hot}$ = 420 °C, $\Delta T \approx 200$ °C) with the crystals placed at the hot end for 16 hours before cooling to room temperature. This process effectively removed the excess tellurium flux, yielding high-quality $MoTe_2$ crystals.

## Device fabrications

The triple-gated $tMoTe_2$ devices were fabricated following procedures similar to those reported previously [9,13,16]. Atomically thin flakes of graphite, hexagonal boron nitride (hBN), 2H-$MoTe_2$, and 2H-$TaSe_2$ were mechanically exfoliated from bulk crystals. Heterostructure assembly was carried out using a standard dry-transfer technique [57] with polycarbonate (PC) stamps inside a nitrogen-filled glove box to minimize sample degradation. The resulting stack, from top to bottom, consisted of a graphite top gate, a 10–30 nm hBN as the top dielectric layer, two pieces of few-layer $TaSe_2$ serving as electrical contacts to $MoTe_2$, twisted bilayer $MoTe_2$ (with the twisted angle controlled by a mechanical rotator during stacking), followed by a 10–30 nm hBN bottom dielectric layer and a graphite bottom gate. The assembled stack was released onto a Si/$SiO_2$ substrate with prepatterned Ti/Au/Pt (3 nm/10 nm/2 nm) electrodes. The entire stack was subsequently annealed at 300 °C for approximately 1 hour in Ar/$H_2$ forming gas. Device patterning into a Hall-bar geometry was then carried out using standard electron-beam lithography and reactive ion etching ($CHF_3$/$O_2$) processes.

## Transport Measurements

Electrical transport measurements above 1.5 K were mainly performed in two closed-cycle $^4$He cryostats equipped with superconducting magnets. Measurements below 1.5 K were mainly performed in a cryogen free dilution refrigerator (Q-one, Q-400) equipped with a 9 T superconducting magnet. The nominal base temperature is about 10 mK. Each fridge line has a sliver epoxy filter and multiple stage RC- filters at low temperature. We performed the electrical transport measurements by using the standard low-frequency lock-in techniques. The bias current is limited within 3 nA to avoid sample heating and disturbance of fragile quantum states.

The longitudinal resistivity $\rho_{xx}$ is obtained from the longitudinal resistance $R_{xx}$ by $\rho_{xx} = R_{xx}\frac{W}{L}$, where $W$ denotes the width of the Hall bar width and $L$ represents the distance between voltage probes. The Hall resistivity $\rho_{xy}$ is equivalent to the measured Hall resistance $R_{xy}$ in two-dimensional case.

**Calibration of twist angles**

The device geometry enables independently control of the carrier density $\left(n = \frac{c_t V_t + c_b V_b}{e} + n_0\right)$ and the vertical electric displacement field $\left(D = \frac{c_t V_t - c_b V_b}{2\varepsilon_0} + D_0\right)$ in $tMoTe_2$ through the application of gate voltages on top and bottom graphite gates, $V_t$ and $V_b$. Here, $\varepsilon_0$, $c_t$, $c_b$, $n_0$ and $D_0$ denote the vacuum permittivity, geometric capacitance of the top graphite gate, geometric capacitance of the bottom graphite gate, intrinsic doping and the built-in electric field, respectively. The values of $c_t$ and $c_b$ are calibrated through quantum oscillations measured under a perpendicular magnetic field $B$ (Extended Data Fig. 1). The moiré hole filling factor $\nu_h$ is determined from a sequence of correlated and/or topological quantum phases with prominent $\rho_{xx}$ and $\rho_{xy}$ features, controlled by the $V_t$ and $V_b$. The carrier density corresponding to $\nu_h$ =1 is defined as the moiré density $n_M$, representing the carrier density required to fill one hole per moiré unit cell. The twist angle $\theta$ and the moiré wavelength $a_M$ are then obtained from $n_M = \frac{2}{\sqrt{3}a_M^2}$ and $a_M = \frac{a}{\sqrt{2(1-\cos\theta)}}$, where $a$ =0.352 nm is the lattice constant of monolayer $MoTe_2$ [58]. The values of $\theta$, $a_M$, $n_M$, for device 1-15 are summarized in the table shown in Extended Data Table 1.

**Estimation of $\xi$ and $l_m$**

The superconducting coherence length $\xi$ is estimated by fitting the temperature dependence of the critical perpendicular magnetic field $B_c$ near $T_c$ using Ginzburg–Landau relation $B_c \approx \frac{\Phi_0}{2\pi\xi^2}\left(1 - \frac{T}{T_c}\right)$, as shown in the inset of Fig. 4f. Here, $\Phi_0 = h/2e$ is the flux quantum, where $h$ and $e$ denote the Planck constant and elementary charge, respectively. The fitted slope yields a coherence length $\xi \approx 44$ nm.

The mean free path $l_m$ is estimated using the two-dimensional Drude model, $l_m = \frac{h}{e^2 k_\mathrm{F} \rho_\mathrm{xx}}$, where $k_\mathrm{F} = \sqrt{2\pi n}$ is the Fermi wave vector assuming a degeneracy of 2. This estimation yields a mean free path $l_m \approx 120$ nm.

## Continuum Model Calculations for tMoTe2

The electronic band structures of tMoTe2 were calculated using a continuum model. Focusing on the low-energy physics originating from the $K$ and $K'$ valleys of the constituent MoTe2 monolayers, the continuum Hamiltonian $\hat{H}_s$ for spin valley $s$ ($s = \pm 1$ for spin up/down, locked to $K/K'$ valleys respectively due to strong Ising-type spin-orbit coupling) is given by:

$$\hat{H}_s = \begin{bmatrix} -\frac{(\hat{\boldsymbol{k}} - \boldsymbol{K}_t)^2}{2m^*} + \Delta_+(\boldsymbol{r}) + \frac{\Delta_D}{2} & \Delta_T(\boldsymbol{r}) \\ \Delta_T^\dagger(\boldsymbol{r}) & -\frac{(\hat{\boldsymbol{k}} - \boldsymbol{K}_b)^2}{2m^*} + \Delta_-(\boldsymbol{r}) - \frac{\Delta_D}{2} \end{bmatrix} \quad (1)$$

with the moiré potentials and interlayer tunneling defined as:

$$\begin{aligned} \Delta_\pm(\boldsymbol{r}) &= 2V_1 \sum_{i=1,3,5} \cos(\boldsymbol{g}_i^1 \cdot \boldsymbol{r} \pm \phi_1) + 2V_2 \sum_{i=1,3,5} \cos(\boldsymbol{g}_i^2 \cdot \boldsymbol{r}) \\ \Delta_T(\boldsymbol{r}) &= w_1 \sum_{i=1,2,3} e^{-i\boldsymbol{q}_i^1 \cdot \boldsymbol{r}} + w_2 \sum_{i=1,2,3} e^{-i\boldsymbol{q}_i^2 \cdot \boldsymbol{r}} \end{aligned} \quad (2)$$

where $\hat{\boldsymbol{k}}$ is the continuous momentum operator, and $\boldsymbol{K}_t$, $\boldsymbol{K}_b$ are the respective zone-corner momenta for the top and bottom monolayer, respectively. The intralayer moiré potential profiles are given by $\Delta_\pm(\boldsymbol{r})$, while $\Delta_T(\boldsymbol{r})$ represents the spatially modulated interlayer electronic hybridization. The momentum transfers $\boldsymbol{g}_i^1$ and $\boldsymbol{g}_i^2$ correspond to intralayer scattering up to the second nearest-neighbor shells in momentum space, whereas $\boldsymbol{q}_i^1$ and $\boldsymbol{q}_i^2$ dictate the interlayer scattering vectors. The effect of the applied vertical displacement electric field $D$ is incorporated via the interlayer potential difference term $\Delta_D = eDd/\varepsilon_{\mathrm{MoTe_2}}$, where $d \approx 0.7$ nm is the physical interlayer distance and $\varepsilon_{\mathrm{MoTe_2}} \approx 10$ accounts for the background dielectric constant characterizing the tMoTe2 environment. The continuum parameters employed in our calculation are from first-principle calculation with density-dependent vdW corrections [59]: $m^* = 0.62 m_e$, $V_1 = 10.3$ meV, $V_2 = 2.9$ meV, $w_1 = -7.8$ meV, $w_2 = 6.9$ meV, and $\phi_1 = -75°$. The reciprocal vectors are scaled by the moiré lattice constant $L_m = a/(2\sin(\theta/2))$, where $a = 3.52$ Å is the lattice constant of monolayer MoTe2.

To obtain the single-particle DOS map, the Hamiltonian was numerically diagonalized in the plane-wave basis truncated at a sufficient cutoff (the 5th moiré momentum shell).

For high-resolution DOS mapping, the moiré Brillouin zone was sampled using a dense $N_k \times N_k$ Monkhorst-Pack mesh with $N_k = 500$ (i.e., $2.5 \times 10^5$ $\boldsymbol{k}$-points per valley).

The total DOS $DOS(\varepsilon)$ was evaluated by integrating over the full Brillouin zone for both valleys:

$$DOS(\varepsilon) = \frac{1}{\Omega_m} \sum_{s,n,\boldsymbol{k}} \frac{1}{\sigma\sqrt{2\pi}} \exp\left(-\frac{(\varepsilon - \varepsilon_{n,s}(\boldsymbol{k}))^2}{2\sigma^2}\right)$$

where $\Omega_m = \sqrt{3}L_m^2/2$ is the moiré unit cell area, $n$ is the band index, and $\sigma$ represents the artificial Gaussian broadening introduced for numerical integration. To cleanly resolve the sharp van Hove singularities characteristic of the flat bands, an adaptive broadening scheme was employed, setting $\sigma \approx w/3N_k$ where $w$ is the bandwidth of the first moiré bands. The integration mapped the energy states into $\nu_h$ to facilitate direct comparison with transport measurements.

**Comparison between $tMoTe_2$ and $tWSe_2$**

In $tWSe_2$, the superconducting region evolves relatively continuously with $\theta$ and $D$-field [21-24]. Near $\theta \approx 3.5°$, superconductivity emerges close to zero displacement field; with increasing $\theta$, it persists but generally requires a progressively larger $D$-field, up to $\theta \approx 5°$. Across this regime, superconductivity is typically found near $\nu_h = 1$ over a broad range of $\theta$ and is generally associated with Fermi surface reconstructions and a nearby intervalley-coherent antiferromagnetic state. We believe the observed superconductivity in relatively large-angle ($\theta > 5.5°$) $tMoTe_2$ shares a similar pairing mechanism as $\theta \approx 3.5°$-$5.5°$ $tWSe_2$.

The microscopic origin of superconductivity in $tWSe_2$ remains under active debate, with existing theories approaching the problem from several different perspectives. Weak-to-intermediate-coupling theories attribute pairing to correlation-enhanced electronic fluctuations, with spin-valley or intervalley-coherent antiferromagnetic fluctuations proposed to mediate unconventional pairing and stabilize nematic, chiral, mixed-parity or topological superconducting states [47-51]. Complementary studies emphasize quantum-critical fluctuations promoted by nontrivial band topology or more general boson-mediated intervalley pairing mechanisms [60–62]. In contrast, strong-coupling theories based on Hubbard- or *t*-*J*-*U* models highlight Mott physics, exchange interactions and correlation-induced renormalization [63–66] as central to superconductivity. In addition, van Hove singularities have also been identified as an important ingredient for enhancing superconductivity in the presence of electronic correlations [47,48,50,67]. Taken together, these theories suggest that superconductivity in $tWSe_2$ may arise from an interplay of van Hove physics, electronic correlations and associated collective fluctuations, although neither the microscopic pairing mechanism nor the pairing symmetry has yet been conclusively established.

In contrast, superconductivity has not yet been observed in $tMoTe_2$ in the intermediate-angle regime $\theta \approx 4°–5.5°$. We tentatively attribute this difference to the stronger effective correlations and more pronounced competition among ordered states in $tMoTe_2$. Compared with $tMoTe_2$, the wider moiré mini bands of $tWSe_2$ reduce the relative interaction strength and may suppress ferromagnetic and Chern-insulating tendencies, allowing antiferromagnetic correlations and their associated fluctuations to develop over a wider parameter range. In intermediate-angle $tMoTe_2$, by contrast, competing ferromagnetic, antiferromagnetic, and topological tendencies may prevent a comparable superconducting phase from becoming established. Consistent with this picture, superconductivity has also not been observed so far in $tWSe_2$ devices with twist angles below ~3.5°, where similar competition between superconducting, ferromagnetic, and topological tendencies may inhibit the emergence of valley-degenerate superconductivity. The detailed evolution of magnetic correlations with twist angle and the microscopic ingredient responsible for the onset of superconductivity remain open questions. Resolving these possibilities will require future measurements using probes that are more directly sensitive to the underlying magnetic order. We also note that the transport phase diagram of device 11 at $\theta = 5.18°$ already closely resembles those observed in the 5.67° $tMoTe_2$ (device 13) and in $tWSe_2$ devices near ~4°. We have therefore performed careful transport measurements on this device at dilution-refrigerator temperatures (Extended Data Fig. 1). Despite its high sample quality, we did not observe superconductivity down to our base temperature. We consider two possible explanations for the absence of superconductivity in device 11. One is that superconductivity is present but remains inaccessible because of the insufficient electronic temperature and/or sample quality. The other is that superconductivity is intrinsically absent in $tMoTe_2$ at this twist angle, in contrast to $tWSe_2$, where superconductivity has been observed over a broad range of $\theta$ in the valley-degenerate regime.

On the other hand, in $tMoTe_2$ with a narrow twist angle range of $\theta \approx 3.83° - 3.93°$, unconventional superconductivity appears near $\nu_h = 0.71$-$0.76$ around $D = 0$, adjacent to FQAH and RIQAH states, and emerges from an at least partially spin-valley-polarized Fermi surface [16]. It is therefore neither centered near $\nu_h = 1$ at finite $D$-fields nor tied to the single-particle VHS filling. These observations indicate that the small-angle superconducting phase in $tMoTe_2$ is unlikely to be a direct counterpart of the superconductivity in $tWSe_2$ and large-angle ($\theta > 5.5°$) $tMoTe_2$, and should instead have a distinct microscopic origin. One possible interpretation is a spin-polarized chiral superconducting state generated by a Kohn–Luttinger-type mechanism [68,69]. A more distinctive scenario, motivated by the FQAH–superconductivity adjacency observed in $\theta \approx 3.83° - 3.93°$ $tMoTe_2$ devices, is anyon superconductivity, in which condensation of composite bosons obtained by doping a fractional Chern insulator drives pairing [70-73]. Determining the pairing symmetry and microscopic mechanism will require further experimental and theoretical work. The devices studied in this work lie outside the narrow twist-angle window where superconductivity near the FQAH state has been observed.

Although the $\nu_h$-$D$ phase diagrams for $tWSe_2$ and $tMoTe_2$ at relatively large $\theta$ are remarkably similar, their evolution toward smaller twist angles is qualitatively distinct. For example, at $\nu_h = 1$ for $tMoTe_2$, the topologically trivial correlated insulator in the layer-hybridized regime is rapidly suppressed with decreasing $\theta$, giving way to the IQAH states as the system enters the topological regime (Fig. 3i). In contrast, in $tWSe_2$, the trivial correlated insulating phase becomes increasingly robust at smaller twist angles and extends over a much broader $D$-field range [23]. This qualitative contrast, demonstrates that, despite the apparent universality in the large-angle limit, topology and correlation evolve with $\theta$ in markedly different ways in the two systems. To date, fractional Chern insulating states and related physics have not been realized in small-angle $tWSe_2$. Understanding the microscopic origin of this qualitative difference remains an important open question.

## Data availability

Data are available from the corresponding authors upon request. Source data are provided with this paper.

## Acknowledgements

We thank J. Shan, K. F. Mak, and F. Wu for valuable discussions. Device fabrication is supported by the Micro-nano Fabrication Platform of the School of Physics and Astronomy at Shanghai Jiao Tong University.

## Funding Statement

This work is supported by the National Key R&D Program of China (Nos. 2022YFA1405400, 2022YFA1402404, 2022YFA1402702, 2022YFA1403301),

National Natural Science Foundation of China (Nos. 12350403, 92565302, 12621003, 12374045, 12350404, 12204115, 92265102, 12550403, 12674229), the Quantum Science and Technology-National Science and Technology Major Project (Nos. 2021ZD0302600, 2021ZD0302500, 2024ZD0300104, 2025ZD0300500), the Natural Science Foundation of Shanghai (No. 24LZ1401100, 24QA2703700), and the Shanghai Jiao Tong University 2030 Initiative Program. J.G., J.Z., and Yuanbo Zhang acknowledge support from the Fundamental and Interdisciplinary Disciplines Breakthrough Plan of the Ministry of Education of China, (Grant No. JYB2025XDXM120), the Shanghai Municipal Science and Technology Commission (Nos. 23JC1400600 and 2019SHZDZX01), and the China Postdoctoral Science Foundation (Grant No. 2022M720812). S.L. acknowledges support by the Hong Kong RGC (No. 26308524, AoE/P-604/25R), the Hong Kong Collaborative Research Fund (No. C6053-23G), the Guangdong Provincial Quantum Science Strategic Initiative (Nos. GDZX2501003, GDZX2601002), and the Ministry of Science and Technology of the People's Republic of China (No. MOST23SC01). K.W. and T.T. acknowledge support from the CREST (JPMJCR24A5), JST and World Premier International Research Center Initiative (WPI), MEXT, Japan. T.L. acknowledges support from the New Cornerstone Science Foundation through the XPLORER PRIZE and the Asian Young Scientist Fellowship (AYSF). X.L. acknowledges "Shuguang Program" supported by Shanghai Education Development Foundation. F.X. are supported by T.D. Lee scholarship.

## Author contributions

Z.S., F.X., J.L., T.J., K.C. and W.T. fabricated the devices and performed the electrical transport measurements. Z.S., F.X., J.L., T.J., K.C., J.J., X.L. and T.L. analysed the data. Y.J., J.G., J.Z., Y.B.Z. and S.L. grew the bulk $MoTe_2$ crystals. C.X. and Y.Z. performed the band-structure calculations. S.W. and S.J. performed optical characterizations. K.W. and T.T. grew the bulk hBN crystals. T.L. and X.L. supervised the project.

## Competing interests

The authors declare no competing interests.

## Main Figures

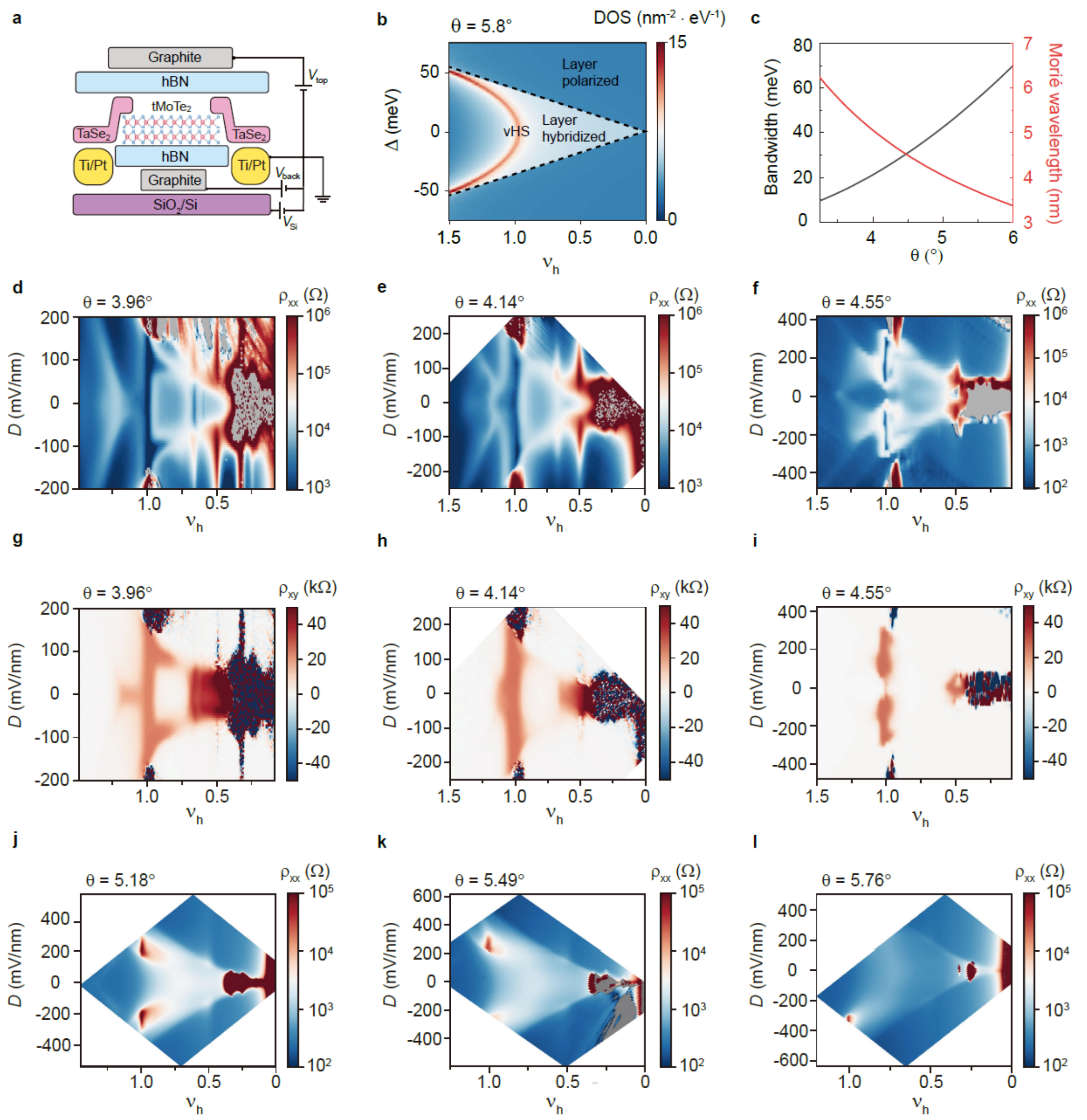

**Fig. 1 | Twist-angle-dependent transport phase diagrams in tMoTe₂. a,** Schematic illustration of the tMoTe₂ device structure used for transport measurements. **b,** Calculated DOS as functions of interlayer potential difference Δ and $\nu_h$ at $\theta$ = 5.8°. The interlayer potential difference can be induced by the applied vertical electric field. **c**, Calculated bandwidth (black) of the first moiré mini band and the moiré wavelength (red) as a function of $\theta$. **d-i,** Longitudinal resistivity $\rho_{xx}$ as functions of $D$ and $\nu_h$ for device 2 (**d**), 4 (**e**), 6 (**f**), 11 (**j**), 12 (**k**), and 15 (**l**), with $\theta$ ranging from 3.96° to 5.76°. Antisymmetrized Hall resistivity $\rho_{xy}$ maps for device 2, 4, and 6 are shown in panel **g**, **h**, and **i**, respectively. Panels **d**,**e**,**g**,**h** were measured under a small perpendicular magnetic field $B$ = 0.1 T, and panels **f**,**i** were measured at $B$ = 0.3 T, whereas panels **j, k,** and **l** were measured at $B$ = 0. Panels **f** and **i** were measured at $T$ = 300 mK, whereas all other maps were measured at $T$ = 1.6 K. Grey regions indicate areas inaccessible to electrical transport measurements due to extremely large sample resistance or poor electrical contacts.

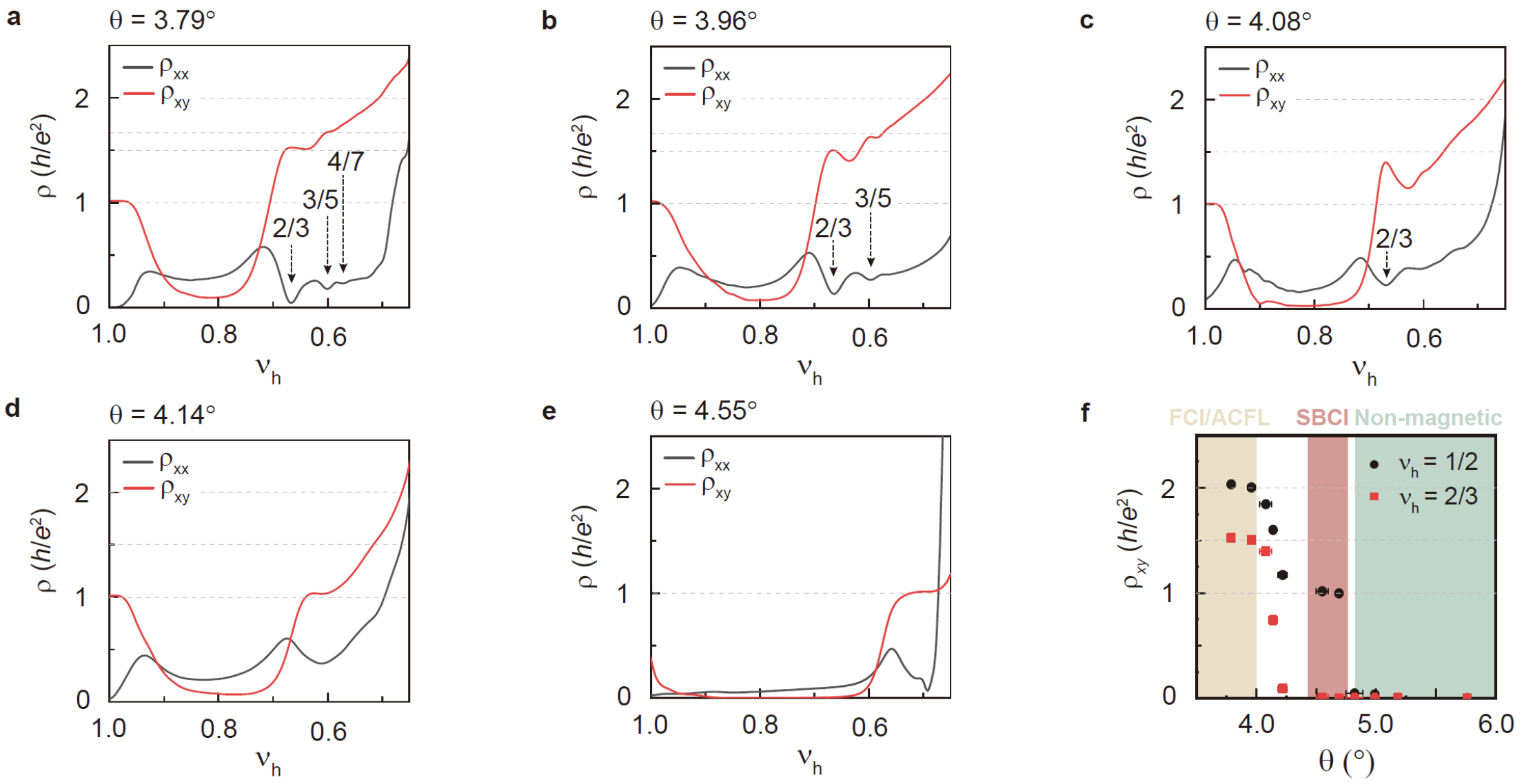

**Fig. 2 | The evolution of fractional topological phases with $\theta$. a-e,** Symmetrized $\rho_{xx}$ and antisymmetrized $\rho_{xy}$ around $D = 0$ as a function of $\nu_h$ for device 1 (**a**), 2 (**b**), 3 (**c**), 4 (**d**), and 6 (**e**), with $\theta$ ranging from 3.79° to 4.55°. Panel **a-d** were measured at $T = 1.6$ K and $B = 0.1$ T, whereas panel **e** was measured at $T = 0.3$ K and $B = 0.3$ T due to the lower Curie temperature at larger $\theta$. **f**, Antisymmetrized $\rho_{xy}$ at $B = 0.1$ T as a function of $\theta$ at $\nu_h = 2/3$ and 1/2. For devices exhibiting ferromagnetism ($\theta < 4.82°$), the $\rho_{xy}$ shown in **f** reflects the anomalous Hall response.

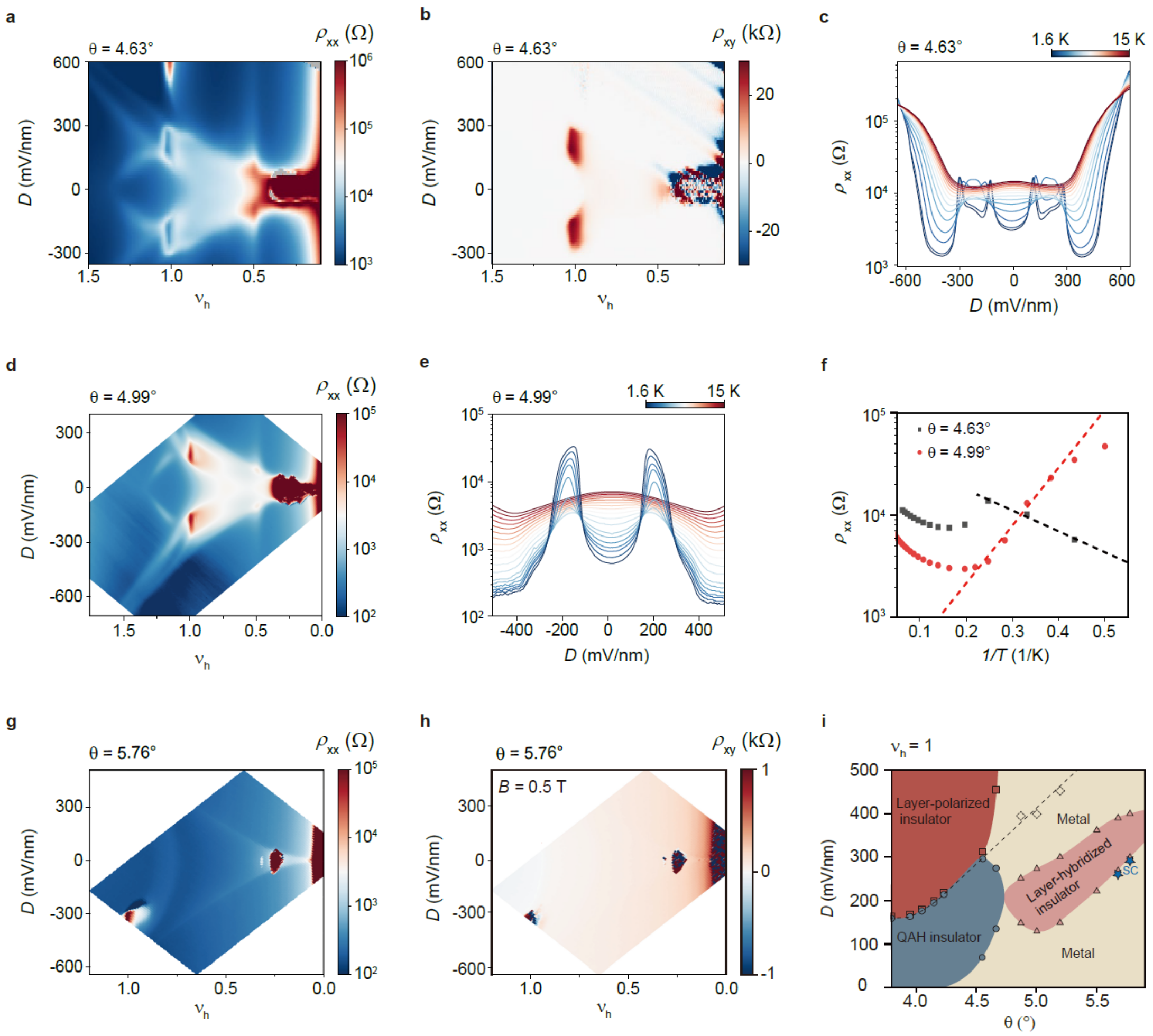

**Fig. 3 | Topological and correlated phases at $\nu_h$ = 1. a,b** Symmetrized $\rho_{xx}$ (**a**) and antisymmetrized $\rho_{xy}$ (**b**) as functions of $\nu_h$ and $D$ for device 7 (4.63°), measured at $T$ = 1.6 K and $B$ = 0.1 T. **c,** Temperature-dependent $\rho_{xx}$ as a function of $D$ at $\nu_h$ =1 for device 7 (4.63°) at $B$ = 0.1 T. **d**,**e**, Same as panels **a** and **c**, respectively, but measured in device 10 (4.99°) at $B$ = 0. **f**, $\rho_{xx}$ as a function of 1/$T$ at $\nu_h$ =1 for device 7 (black dots, measured at $D$ = -225 mV/nm) and device 10 (red dots, measured at $D$ = 190 mV/nm). Both devices exhibit thermally activated behavior at low temperatures, as indicated by the dashed lines. The $\rho_{xx}$ of the trivial correlated insulator increases as the temperature decreases, whereas that of the IQAH insulator exhibits the opposite trend due to the presence of chiral edge states. Thermal activation fitting yields an energy gap of approximately 2 meV for the trivial insulating state and approximately 1 meV for the IQAH insulating state. **g**, $\rho_{xx}$ as functions of $\nu_h$ and $D$ for device 15 (5.76°), measured at $T$ = 10 mK and $B$ = 0. **h**, Antisymmetrized $\rho_{xy}$ as functions of $\nu_h$ and $D$ for device 15 (5.76°), measured at $T$ = 10 mK and $B$ = 0.5 T. **i**, $\theta$-$D$ phase diagram at $\nu_h$ = 1 constructed from experimental data (symbols). Dark red denotes the layer-polarized correlated insulator, dark blue the IQAH insulator, pink the layer-hybridized correlated insulator, yellow the metallic phase, and blue stars represent superconducting phase. The dashed line marks the boundary between the layer-polarized and the layer-hybridized region.

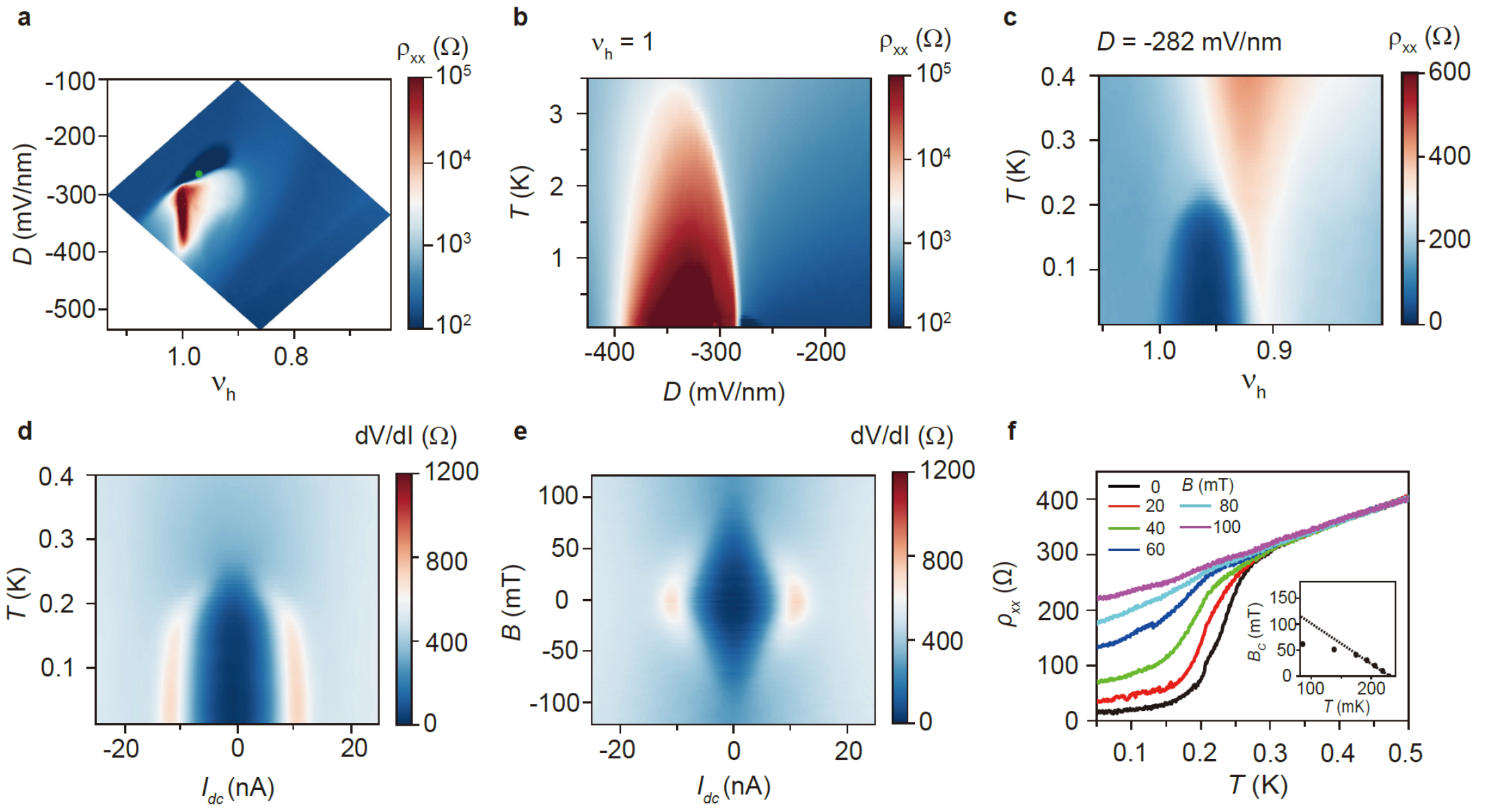

**Fig. 4 | Superconductivity in 5.76° tMoTe₂. a,** Local $\rho_{xx}$ map as functions of $\nu_h$ and $D$ for device 11 (5.76°), measured at $T = 10$ mK and $B = 0$. **b,** $\rho_{xx}$ as functions of $T$ and $D$ at $\nu_h = 1$. **c,** $\rho_{xx}$ as functions of $T$ and $\nu_h$ at $D = -282$ mV/nm. **d**, Differential resistivity d$V$/d$I$ as functions of $I_{dc}$ and $T$ at $B = 0$ for the superconducting state. **e**, The $B$-d$V$/d$I$ –$I_{dc}$ map measured at $T = 10$ mK. **f**, $\rho_{xx}$ as a function of $T$ under different $B$. The inset show the superconducting critical perpendicular magnetic field $B_c$ as a function of $T$. The dashed line represents the liner fit to the data points above 150 mK. Measurements in panels **d-f** were performed at the optimal superconducting condition ($\nu_h = 0.96$, $D = -288$ mV/nm), as marked by the green dot in **a**.

## Extended Data Figures and Tables

| Device | Bulk crystal | Max monolayer Hall mobility ($cm^2$/Vs) * | Twist angle (degree) ** | Moiré wavelength (nm) | Moiré density ($10^{12}$ $cm^{-2}$) |
|---|---|---|---|---|---|
| D1 | HKUST | ~ 40000 | 3.79±0.03 | 5.32±0.04 | 4.08±0.07 |
| D2 | HKUST | ~ 10000 | 3.96±0.03 | 5.09±0.04 | 4.45±0.06 |
| D3 | FDU | ~ 15000 | 4.08±0.05 | 4.94±0.06 | 4.72±0.12 |
| D4 | FDU | ~ 15000 | 4.14±0.04 | 4.87±0.05 | 4.86±0.09 |
| D5 | HKUST | ~ 10000 | 4.22±0.04 | 4.78±0.04 | 5.06±0.09 |
| D6 | FDU | ~ 15000 | 4.55±0.04 | 4.43±0.04 | 5.87±0.10 |
| D7 | FDU | ~ 15000 | 4.63±0.05 | 4.36±0.05 | 6.08±0.13 |
| D8 | FDU | ~ 40000 | 4.69±0.01 | 4.30±0.01 | 6.25±0.03 |
| D9 | HKUST | ~ 9000 | 4.82±0.07 | 4.18±0.06 | 6.59±0.18 |
| D10 | HQ graphene | ~ 5000 | 4.99±0.03 | 4.04±0.03 | 7.07±0.09 |
| D11 | HKUST | ~ 40000 | 5.18±0.01 | 3.90±0.01 | 7.60±0.04 |
| D12 | FDU | ~ 15000 | 5.49±0.02 | 3.67±0.01 | 8.56±0.06 |
| D13 | HKUST | ~ 26000 | 5.67±0.02 | 3.55±0.02 | 9.13±0.06 |
| D14 | FDU | ~ 40000 | 5.69±0.03 | 3.54±0.02 | 9.19±0.09 |
| D15 | HKUST | ~ 40000 | 5.76±0.02 | 3.50±0.01 | 9.42±0.05 |

**Extended Data Table 1. Device summary table.** * The 'Maximum monolayer Hall mobility' refers to the highest Hall mobility measured at $T \approx 1.5$–2 K in monolayer $MoTe_2$ devices fabricated from the same bulk-crystal batch as the corresponding moiré device. This quantity provides a useful proxy for the intrinsic quality of each bulk crystal. Device 10 was fabricated using commercially available CVT-grown crystals from HQ Graphene, whereas all other devices were fabricated using flux-grown crystals from different batches.

**The uncertainties in the extracted twist angle, moiré wavelength, and moiré density include both fitting uncertainty and the variation among neighboring contact pairs [16]. These uncertainties therefore provide a quantitative indicator of moiré inhomogeneity.

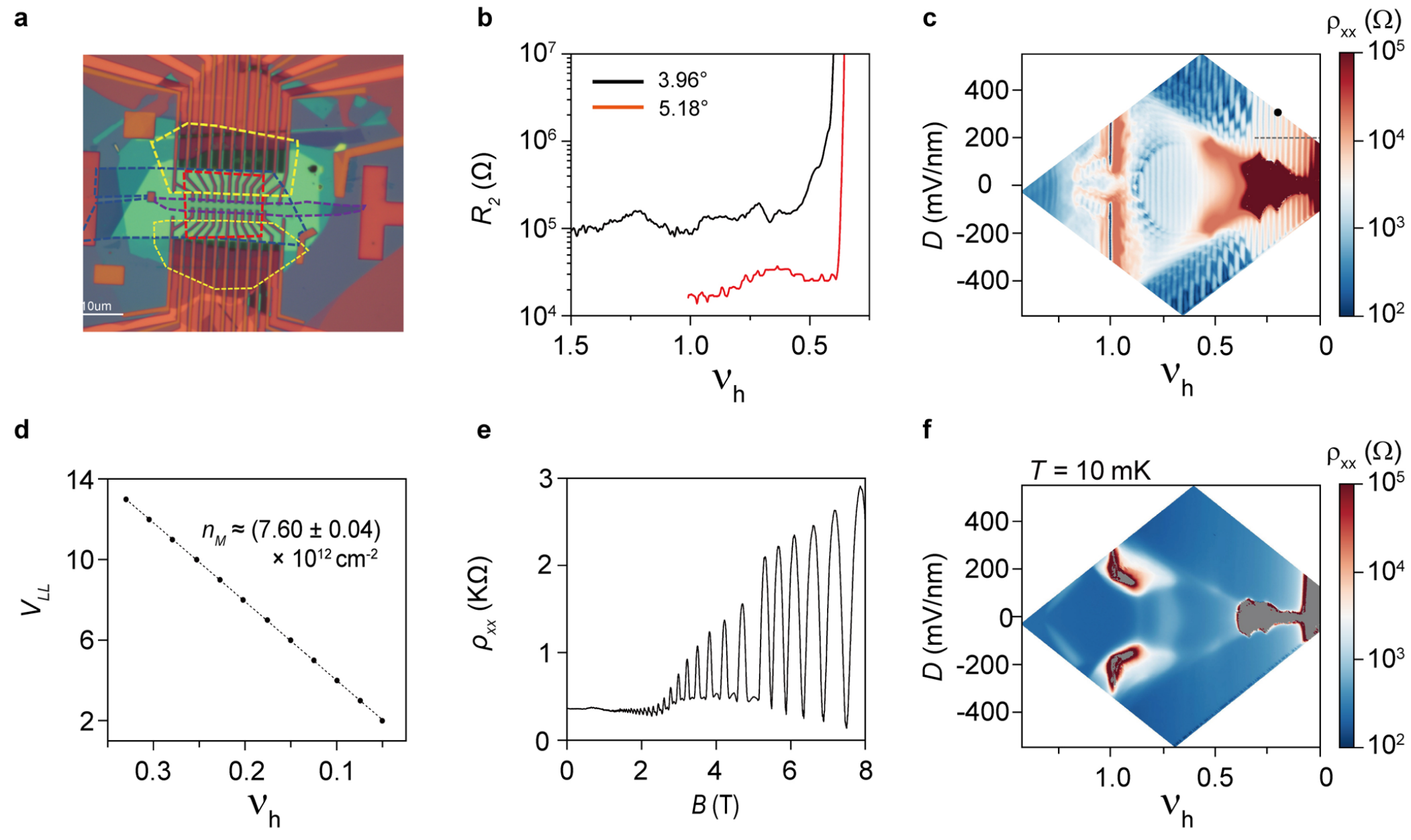

**Extended Data Fig. 1. | Device image and twist angle calibration for device 11 (5.18°). a**, Optical micrograph image of device 11 (5.18°). Yellow dashed lines indicate the $TaSe_2$ contacts. Red dashed lines outline the $tMoTe_2$ region. Blue and purple dashed lines denote the top-gate and bottom-gate regions, respectively. **b**, Two-terminal resistance $R_2$ as a function of $\nu_h$ for device 2 (3.96°) and device 11 (5.18°), shown by the black and red curves, respectively. The measured $R_2$ is primarily limited by contact resistance. In general, devices with larger twist angles exhibit better contacts, possibly due to the less flat moiré bands. **c**, Map of $\rho_{xx}$ as a function of $\nu_h$ and $D$ for device 11 (5.18°) measured at $T$ = 10 mK and $B$ = 8 T. **d**, Linear fit of the Landau-level filling factor $\nu_{LL}$ a function of $\nu_h$ (along the dashed line in **c**), from which the moiré density $n_M$ is extracted. **e**, $\rho_{xx}$ as a function of $B$ at $D$ = 290 mV/nm and $\nu_h$ = 0.25, corresponding to the black dot marked in **c**. The onset of Shubnikov–de Haas oscillations occurs at $B \approx 1.3$ T, reflecting the high quality of the device. **f**, $\rho_{xx}$ as a function of $\nu_h$ and $D$ at $T$ = 10 mK and $B$ = 0 T. No superconducting state is observed in this device. Grey regions indicate areas inaccessible to electrical transport measurements.

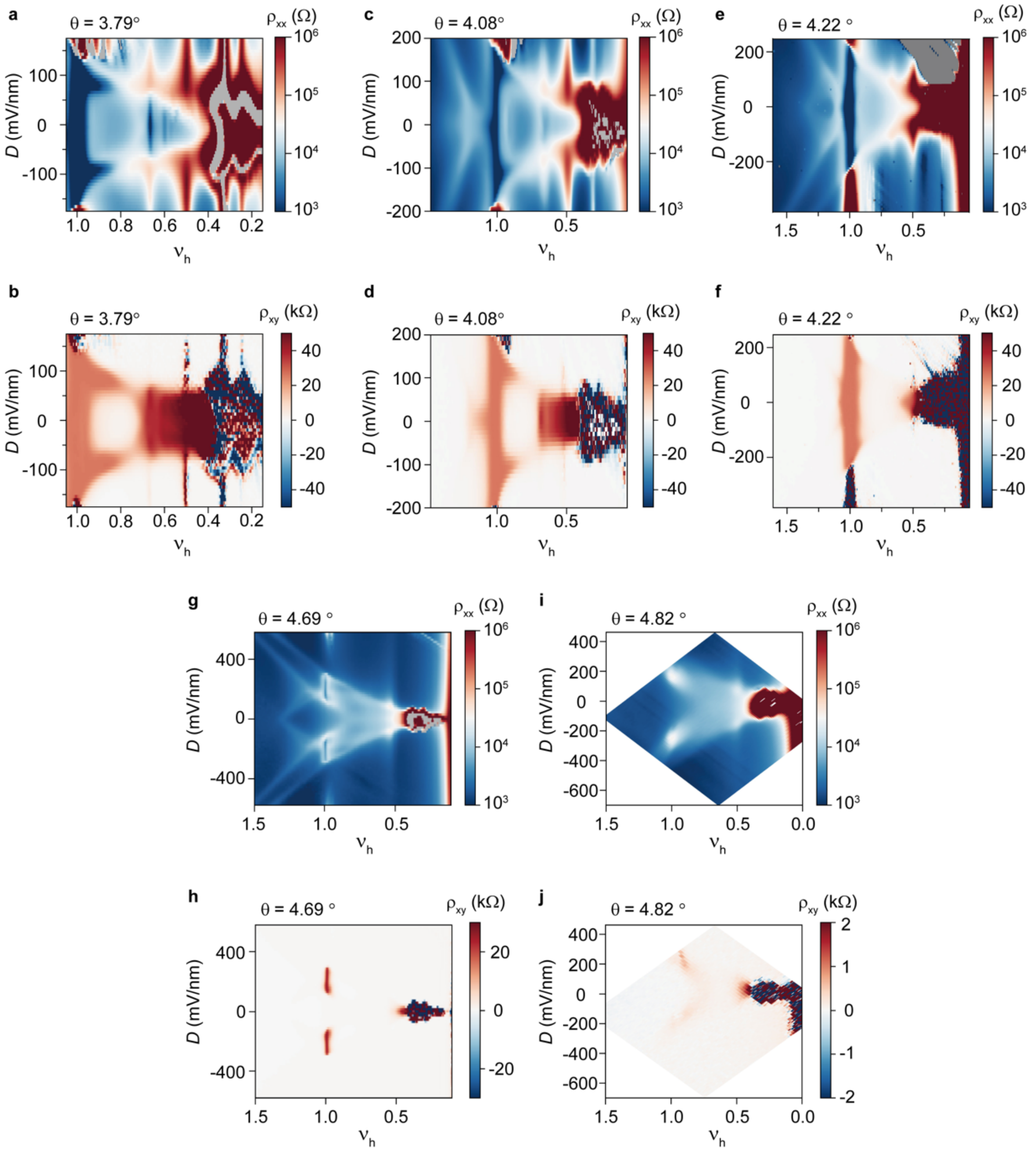

**Extended Data Fig. 2. | Transport phase diagrams of device 1 (3.79°), device 3 (4.08°), device 5 (4.22°), device 8 (4.69°), and device 9 (4.83°). a**,**c**,**e**,**g**,**i**, Symmetrized $\rho_{xx}$ as a function of $D$ and $\nu_h$ for device 1 (**a**), 3 (**c**), 5 (**e**), 8 (**g**), and 9 (**i**). **b**,**d**,**f**,**h**,**j**, Antisymmetrized $\rho_{xy}$ as a function of $D$ and $\nu_h$ for device 1 (**b**), 3 (**d**), 5 (**f**), 8 (**h**), and 9 (**j**). All maps measured at $T = 1.6$ K and with $B = 0.1$ T (**a-f**) or 0.3 T (**g-j**). Device 1 (3.79°) host multiple FQAH states. As the twist angle increases, only IQAH states remain in device 5 (4.22°) and 8 (4.69°), while no quantum anomalous Hall states are observed in device 9 (4.83°). Grey regions in **a**,**c**,**e** indicate areas inaccessible to electrical transport measurements.

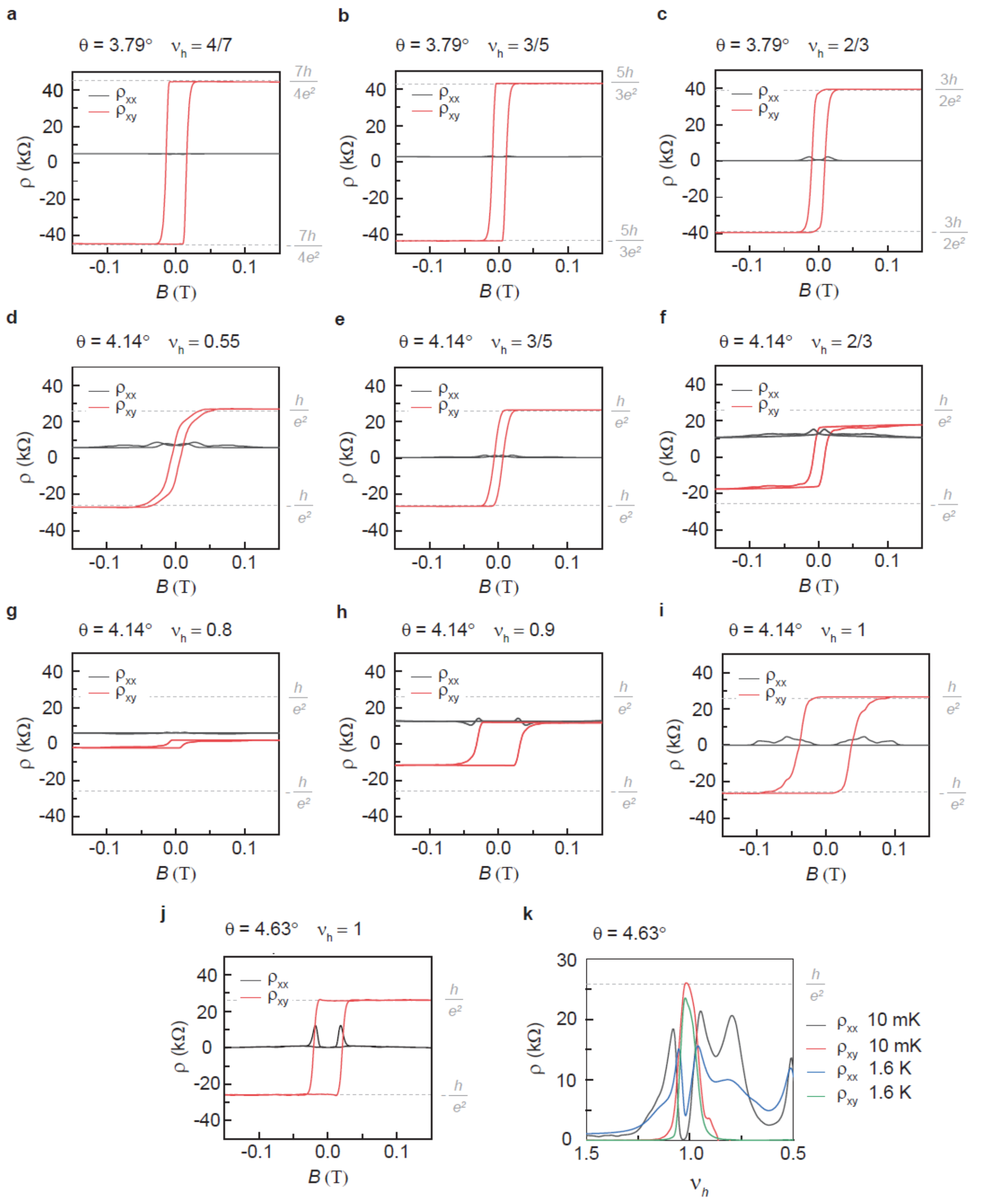

**Extended Data Fig. 3. | Magnetism and Chern insulators in device 1 (3.79°), device 4 (4.14°) and device 7 (4.63°). a-c**, Magnetic hysteresis loop scans at $\nu_h = 4/7$, 3/5, and 2/3, respectively, measured at $D = 0$ in device 1. Clear FQAH effects are observed at these fillings. **d-i**, Magnetic hysteresis loop scans at $\nu_h$ = 0.55, 3/5, 2/3, 0.8, 0.9 and 1, respectively, measured at $D = 0$ in device 4. In this device, robust ferromagnetism persists over the filling range of $\nu_h \approx 1/2$ to 1 at $D = 0$, and IQAH effects are observed around both $\nu_h = 1$ and $\nu_h \approx 0.62$. Panels **a-i** were measured at $T = 600$ mK. **j**, Magnetic hysteresis loop scans measured at $\nu_h = 1$ and $D = 175$ mV/nm in device 7 at $T = 10$ mK. **k**, Symmetrized $\rho_{xx}$ and antisymmetrized $\rho_{xy}$ as a function of $\nu_h$ at $D = 175$ mV/nm measured at $T = 10$ mK (red and black curves) and at $T = 1.6$ K (green and blue curves), the applied $B$-field is 0.1 T. At $T = 10$ mK, $\rho_{xx}$ at the IQAH state nearly vanishes, while at $T = 1.6$ K, the residual $\rho_{xx}$ increases to ~5 kΩ and $\rho_{xy}$ deviates from the quantized value of $h/e^2$.

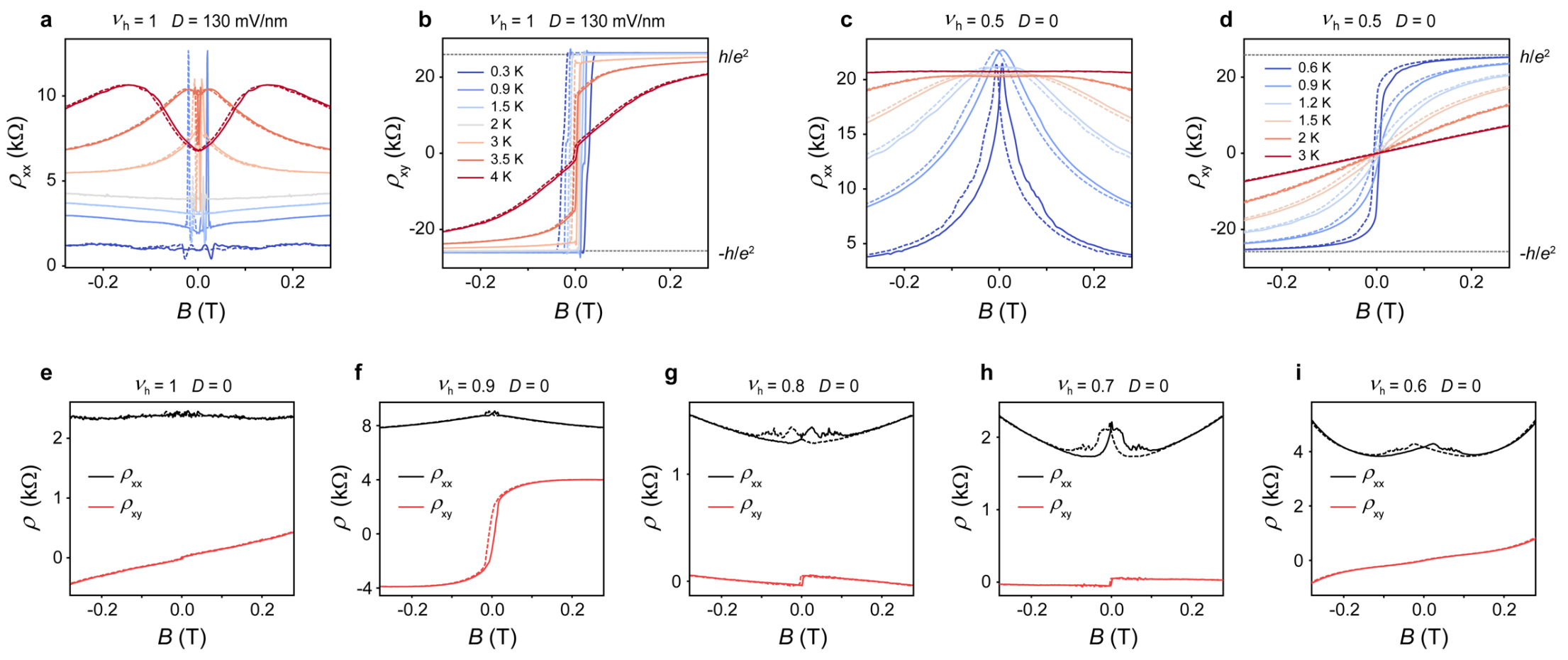

**Extended Data Fig. 4. | Magnetism and Chern insulators in device 6 (4.55°). a**,**b**, Temperature dependence of magnetic hysteresis loop scans of $\rho_{xx}$ (**a**) and $\rho_{xy}$ (**b**) at $\nu_h = 1$ and $D = 130$ mV/nm. Integer quantum anomalous Hall effect is observed at low temperatures and at $B = 0$. The Curie temperature for the ferromagnetism at $\nu_h = 1$ is approximately 4 K. **c**,**d**, The same as **a** and **b**, respectively, but measured at $\nu_h = 1/2$ and $D = 0$. Clear magnetic hysteresis loop is observed at low temperatures, but zero-field quantization is not achieved. A small magnetic field $B \approx 0.2$ T is required to stabilize the SBCI state. The Curie temperature at $\nu_h = 1/2$ is approximately 2 K. **e-i**, Magnetic hysteresis loop scans at $\nu_h = 1$, 0.9, 0.8, 0.7, and 0.6, respectively, measured at $T = 300$ mK and $D = 0$, respectively. Ferromagnetism is absent around $\nu_h = 1$ and 0.6 at $D = 0$.

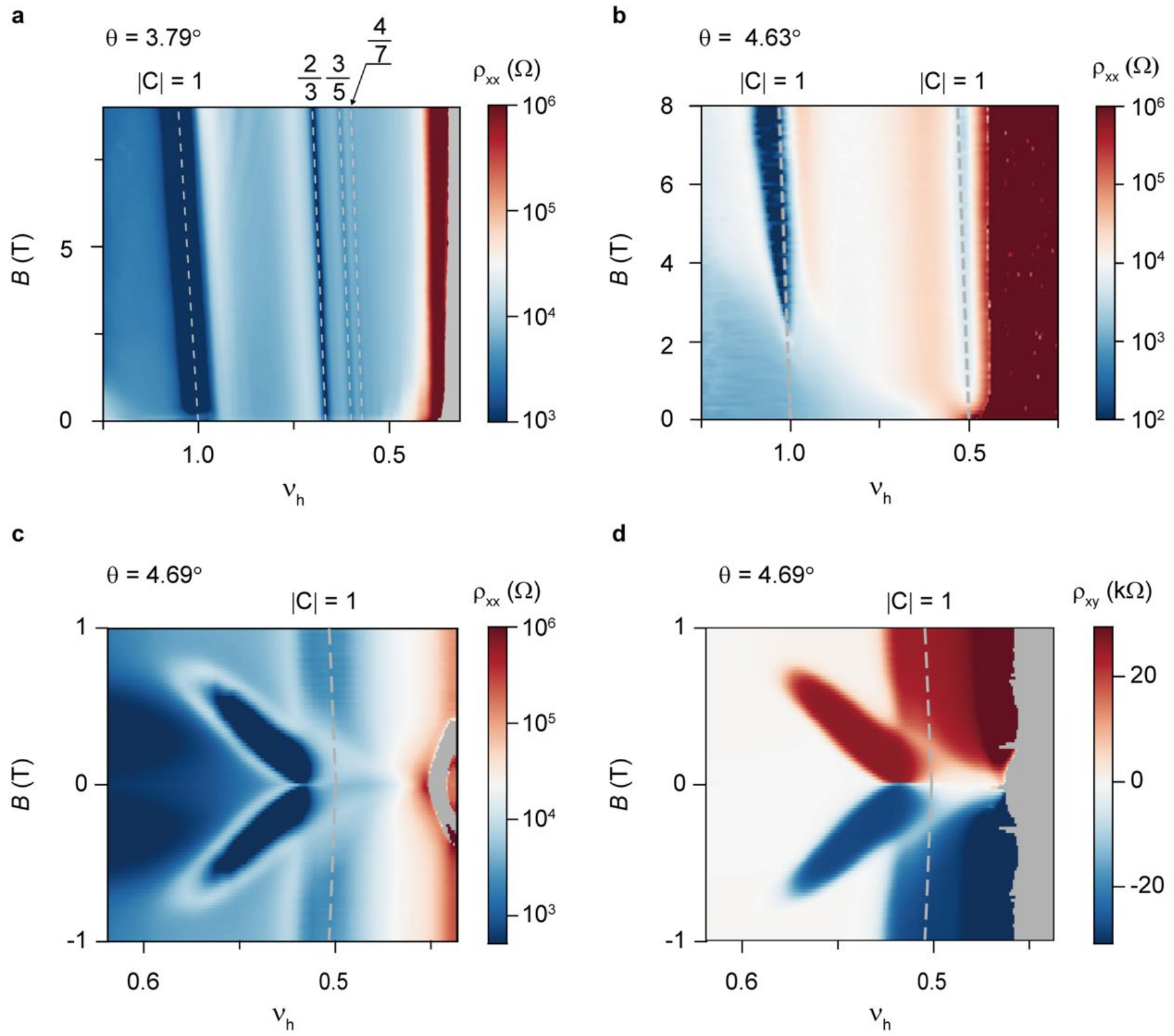

**Extended Data Fig. 5. | Středa formula for integer, fractional, and symmetry-broken Chern insulators. a-b**, $\rho_{xx}$ as a function of $\nu_h$ and $B$ for device 1 (**a**) and device 7 (**b**). Panel **a** was measured at 1.6 K and Panel **b** at 0.3 K. In device 1 (3.79°), the integer and fractional Chern insulators follows the expected dispersions based on the Středa formula $n_M \frac{d\nu_h}{dB} = C\frac{e}{h}$, as indicated by the dashed lines. In device 7 (4.63°), the SBCI states at $\nu_h$ = 1/2 also approximately follows the Streda formula with $C$ = 1, consistent with the measured quantized $\rho_{xy}$ of $h/e^2$. **c-d**, $\rho_{xx}$ (**c**) and $\rho_{xy}$ (**d**) as a function of $\nu_h$ and $B$ for device 8 (4.69°). Grey regions in maps indicate areas inaccessible to electrical transport measurements. In addition to the SBCI state at $\nu_h$ = 1/2, a new state emerges from $\nu_h \approx 0.52$. This state also exhibits a quantized anomalous Hall resistance of $h/e^2$ but does not follow the Středa relation. Interestingly, a closely related state (at $\nu_h \approx 0.53$) was recently reported in a 3.7° twisted mono-bilayer $MoTe_2$ device [74], indicating that both twist-angle tuning and remote-layer engineering can give rise to analogous modifications of the band quantum geometry. We also note a very recent study reporting similar observation in a 4.54° $tMoTe_2$ [75].

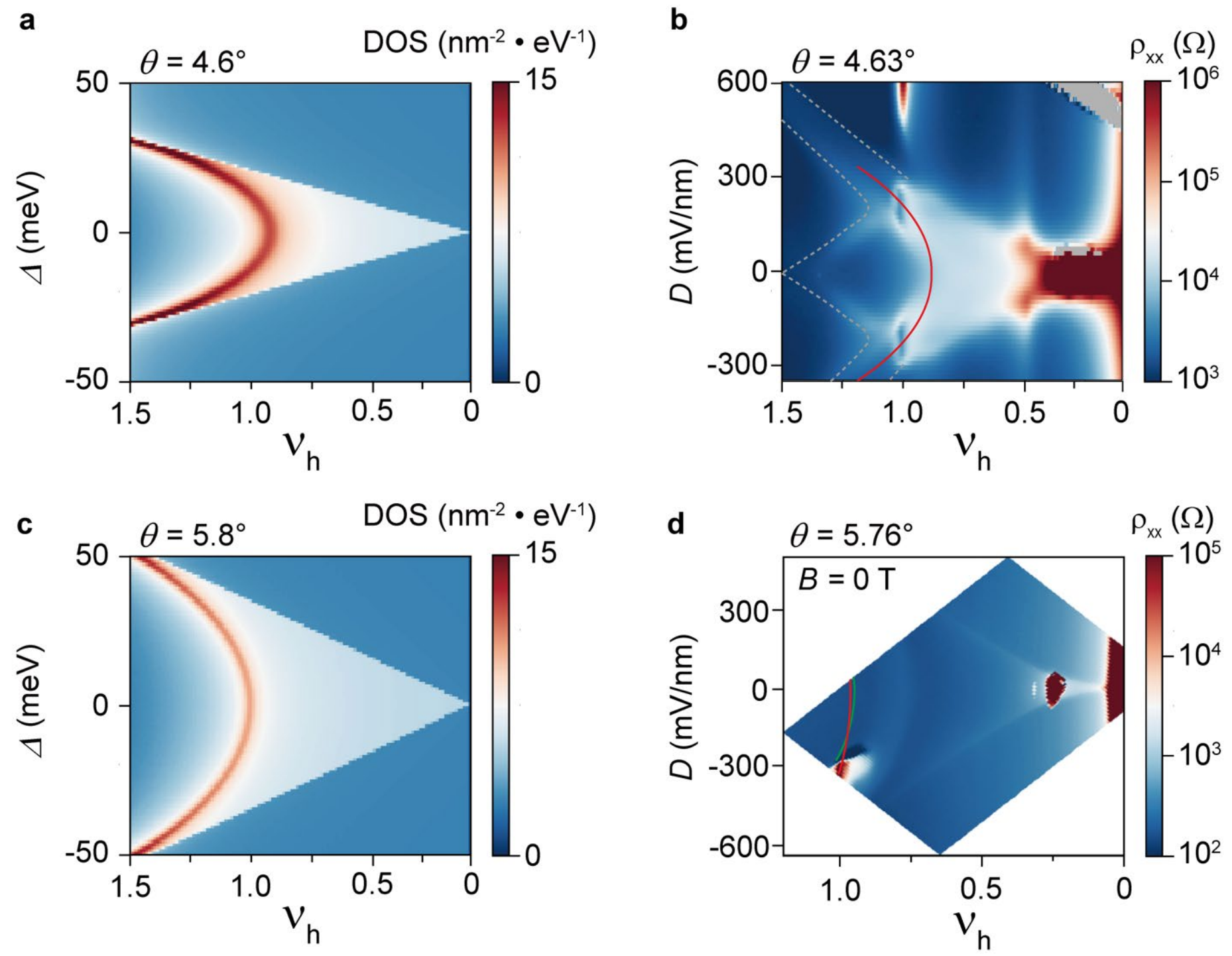

**Extended Data Fig. 6. | Calculated DOS compared with experiments. a,c**, Calculated DOS as a function of interlayer potential difference Δ and $\nu_h$ at $\theta = 4.6°$(**a**) and $\theta = 5.8°$(**c**). **b**, $\rho_{xx}$ as a function of $\nu_h$ and $D$ at $B = 0$ and $T = 1.6$ K for device 7 (4.63°). The red solid line marks the calculated VHS from **a**. We approximately convert Δ into $D$ using $D = E\varepsilon_{\mathrm{hBN}} = \Delta\varepsilon_{\mathrm{MoTe_2}}/(ed)$, where $d$ is the interlayer separation in $MoTe_2$, and $\varepsilon_{\mathrm{hBN}}$, $\varepsilon_{\mathrm{MoTe_2}}$ are the dielectric constant for hBN and $MoTe_2$, respectively (Methods). It illustrates that the emergence of the IQAH state roughly coincides with the $D$-field at which the VHS crosses $\nu_h = 1$. Dashed lines mark the local $\rho_{xx}$ peaks beyond single-particle VHS, which is likely associated with interaction-induced Fermi-surface reconstruction. **d**, Comparison between the calculated VHS (red) and the experimentally extracted VHS (green) from the $\rho_{xy}$-$\nu_h$-$D$ map of device 15 (shown in Fig. 3**h)**, shown on the $\rho_{xx}$-$\nu_h$-$D$ map at $B = 0$ T and $T = 10$ mK. The calculated VHS across the superconducting and correlated insulating states, also agree reasonably well with the experimentally extracted VHS.

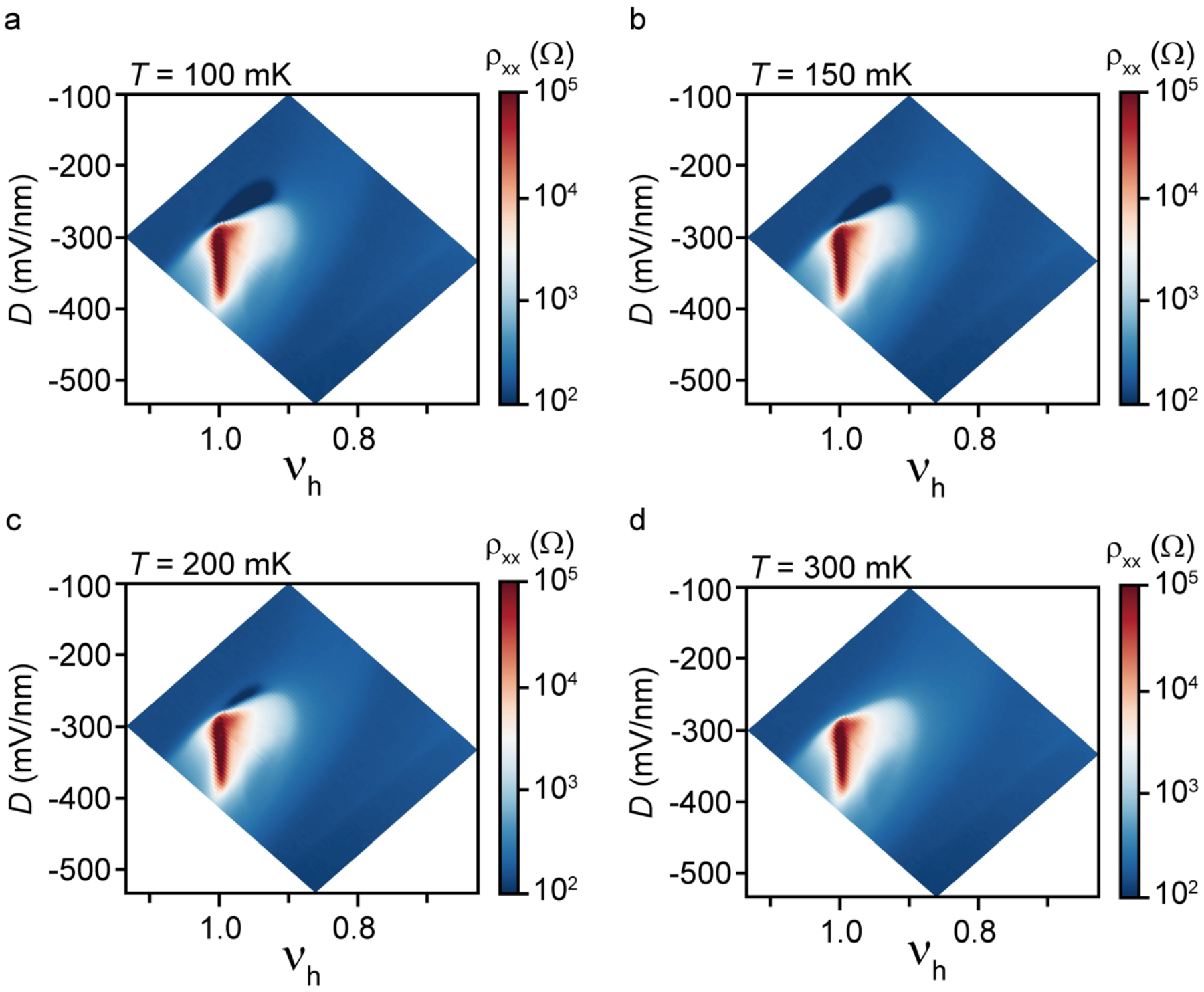

**Extended Data Fig. 7. | *T*-dependent $\rho_{xx}$ maps in device 15 (5.76°). a-d**, Maps of $\rho_{xx}$ at zero magnetic field as a function of $\nu_h$ and $D$, measured at $T$ = 100 mK, 150 mK, 200 mK, 300 mK, respectively. Observed superconductivity is most robust at fillings slightly below $\nu_h$ = 1, adjacent to the $\nu_h$ = 1 correlated insulator.

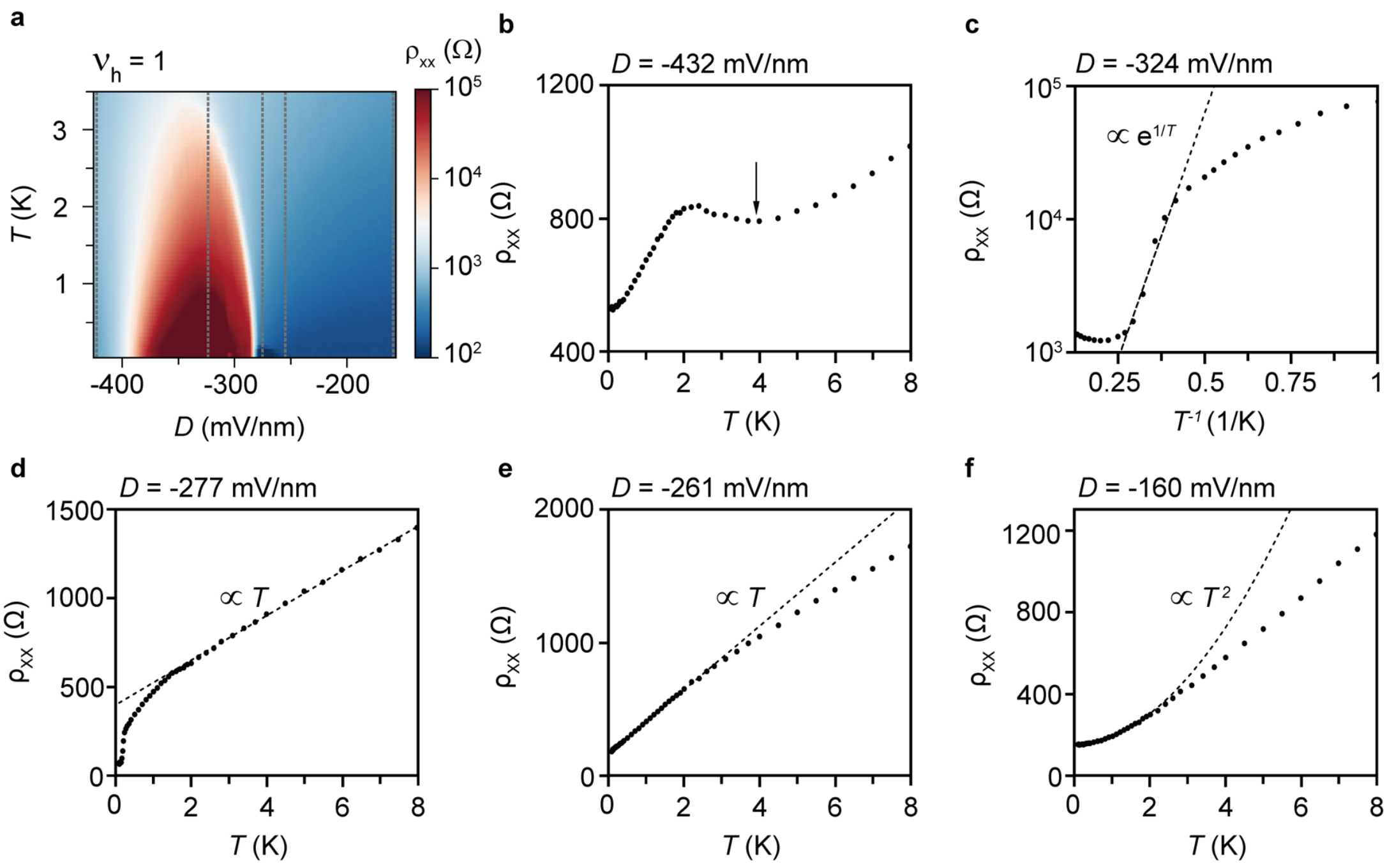

**Extended Data Fig. 8. | *D*-dependent transport behaviors at $\nu_h$ = 1 in device 15 (5.76°). a**, $\rho_{xx}$ as a function of *T* and *D* at zero magnetic field, the same dataset as shown in Fig. 4b. Line cuts shown in **b-f** are indicated by the vertical dashed lines. **b**, $\rho_{xx}$ as a function of *T* at *D* = -432 mV/nm. A non-monotonic temperature dependence is observed, with a local $\rho_{xx}$ minimum around 4 K. Similar behaviors have been reported in $tWSe_2$ [23], where the local $\rho_{xx}$ minimum roughly indicate the Néel temperature of the valley-coherent antiferromagnetic order. **c**, $\rho_{xx}$ as a function of $1/T$ at *D* = -324 mV/nm. Thermal activation behavior corresponding to a correlated insulator gap is observed. The fit yields a gap of ~ 2.5 meV. **d-f**, $\rho_{xx}$ as a function of *T* at *D* = -277, -261, and -160 mV/nm, respectively. Strange-metal behaviors with linear-in-*T* resistivity are observed in the superconducting normal state (**d**) and adjacent to the superconducting region (**e**). Away from the superconducting region at smaller *D*, Fermi liquid behaviors are observed at low temperatures (**f**).

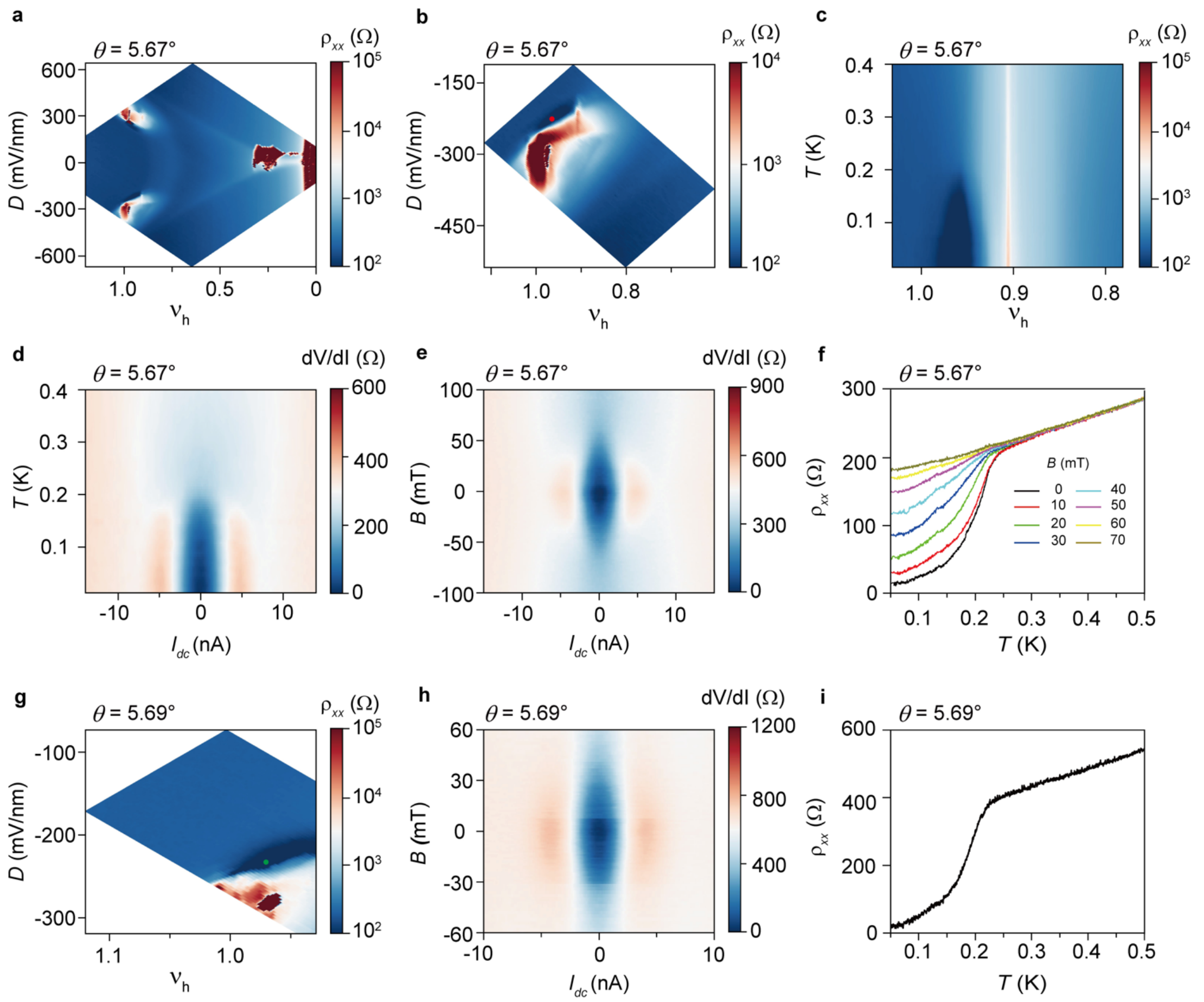

**Extended Data Fig. 9. | Superconducting states in device 13 (5.67°) and device 14 (5.69°).** **a-b,** $\rho_{xx}$ map as a function of $\nu_h$ and $D$ for device 13 (5.67°), measured at $T$ = 10 mK and $B$ = 0. **c**, $\rho_{xx}$ as a function of $T$ and $\nu_h$ at $D$ = -220 mV/nm. **d**, Differential resistivity d$V$/d$I$ as a function of $I_{dc}$ and $T$ at $B$ = 0 for the superconducting state. **e**, The $B$-d$V$/d$I$ –$I_{dc}$ map measured at $T$ = 10 mK. **f**, $\rho_{xx}$ as a function of $T$ under different $B$. Measurements in panels **d-f** were performed at the optimal superconducting condition for device 13 ($\nu_h$ = 0.965, $D$ = -227 mV/nm), as marked by the red dot in **b**. **g,** Local $\rho_{xx}$ map as a function of $\nu_h$ and $D$ for device 14 (5.69°), measured at $T$ = 10 mK and $B$ = 0. The range shown in **g** is limited by gate leakage, which prevents us from accessing a more complete phase diagram in this device. **h**, The $B$-d$V$/d$I$ –$I_{dc}$ map measured at $T$ = 10 mK. **i**, $\rho_{xx}$ as a function of $T$ at $B$ = 0. Measurements in panels **h**,**i** were performed at the optimal superconducting condition for device 14 ($\nu_h$ = 0.97, $D$ = -230 mV/nm), as marked by the green dot in **g**. The optimal superconducting transition temperature $T_c$ (determined by 50% $R_n$) is approximately 175 mK in device 13 and 165 mK in device 14, respectively, which is slightly lower than that observed in device 15 with a larger twist angle of 5.76°.

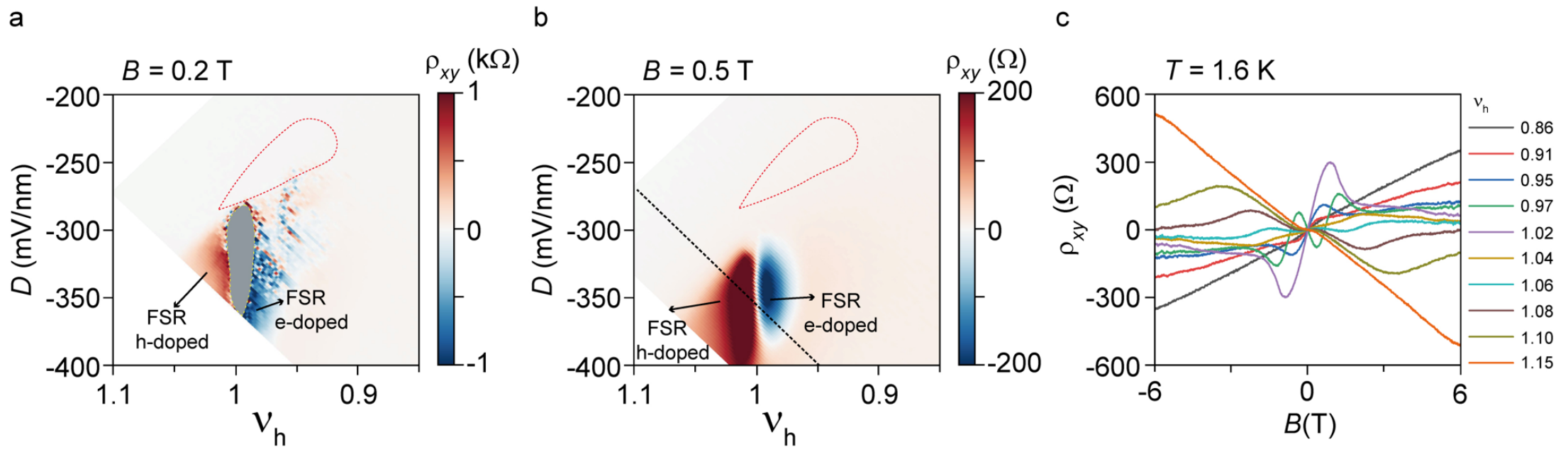

**Extended Data Fig. 10. | Fermi surface reconstructions near the superconducting state in device 15 (5.76°). a-b**, Antisymmetrized $\rho_{xy}$ as a function of $D$ and $\nu_h$ measured at $B$ = 0.2 T and $T$ = 10 mK (**a**), and at $B$ = 0.5 T and $T$ = 1.6 K (**b**). The red dashed lines delineate the superconducting region based on the $\rho_{xx}$ map shown in Fig. 4a. Fermi Surface reconstructions (FSR) are observed near the correlated insulator at $\nu_h$ = 1, resembling those recently reported in $tWSe_2$ [23,24]. **c**, $\rho_{xy}$ as a function of $B$ at different $\nu_h$ following the black dashed line shown in **b**, measured at $T$ = 1.6 K.